\documentclass[12pt]{article}
\usepackage{graphicx} 
\usepackage{booktabs}
\usepackage{amsmath, amssymb, bm}
\usepackage{comment}
\usepackage{float}
\usepackage{tabularx}
\usepackage{amsfonts}
\usepackage[super,sort&compress]{natbib}
\usepackage{xcolor}
\usepackage{url}
\usepackage{multirow}
\usepackage{pdflscape}
\usepackage{geometry}
\usepackage{soul}
\usepackage{adjustbox}
\usepackage{placeins}
\usepackage{authblk}
\usepackage{microtype}
\usepackage[hidelinks]{hyperref}
\newcommand{\doi}[1]{\url{https://doi.org/#1}}

\newcommand{\sh}[1]{\textcolor{black}{#1}}

\newcommand{\ManuscriptTitle}{Weighting Scheme and Construction in IPCW Survival Estimation}
\newcommand{\EqualContribution}{These authors contributed equally to this work and are listed in alphabetical order.}
\newcommand{\MoffittAffiliation}{Department of Biostatistics and Bioinformatics, H. Lee Moffitt Cancer Center and Research Institute, Tampa, FL, USA}
\newcommand{\StatisticsAffiliation}{Department of Statistics and Data Science, Northwestern University, Evanston, IL, USA}
\newcommand{\PreventiveAffiliation}{Department of Preventive Medicine, Division of Biostatistics and Informatics, Northwestern University, Chicago, IL, USA}
\newcommand{\UCSDaffiliation}{Division of Biostatistics and Bioinformatics, University of California San Diego, San Diego, CA, USA}
\newcommand{\CorrespondingEmail}{\href{mailto:ruohui.chen@moffitt.org}{ruohui.chen@moffitt.org}}

\newcommand{\AuthorContributions}{R.C. and S.L. contributed equally to this work. R.C. and S.L. conceived and designed the study. S.L. derived the theoretical results and conducted the simulations and data analysis. S.Z. reviewed the code and results for the simulations and data analysis. R.C. and S.L. drafted the manuscript. R.C. provided supervision. All authors critically revised the manuscript and approved the final version.}

\newcommand{\FundingStatement}{This work was supported in part by the National Cancer Institute of the National Institutes of Health under Award Number P30CA076292, and by a Moffitt Cancer Center Biostatistics and Bioinformatics Methods Development Award. The content is solely the responsibility of the authors and does not necessarily represent the official views of the National Institutes of Health.}

\newcommand{\CompetingInterests}{The authors declare no conflicts of interest.}

\newcommand{\EthicsApproval}{This study is a secondary analysis of the publicly available WIHS dataset distributed with the R package \texttt{randomForestSRC}. No additional ethics approval was required.}

\newcommand{\ConsentParticipation}{No additional informed consent was required for this secondary analysis of publicly available data.}

\newcommand{\ConsentPublication}{Not applicable.}

\newcommand{\AIUseDeclaration}{The authors used ChatGPT, Grammarly, and Copilot to improve the language and readability of the manuscript. After using these tools, the authors edited and verified the content, and take full responsibility for the final production.}

\title{\ManuscriptTitle}
\author[1*,$\dagger$]{Ruohui Chen}
\author[2*]{Shuhan Liu}
\author[3]{Scott Zuo}
\author[3]{Lihui Zhao}
\author[3]{Hui Zhang}
\author[4]{Xin Tu}
\author[1]{Dung-Tsa Chen}
\author[1]{Qianxing Mo}
\affil[1]{\MoffittAffiliation}
\affil[2]{\StatisticsAffiliation}
\affil[3]{\PreventiveAffiliation}
\affil[4]{\UCSDaffiliation}
\affil[*]{\EqualContribution}
\affil[$\dagger$]{Corresponding author: \CorrespondingEmail}
\date{}

\begin{document}

\maketitle

\section*{Abstract}
Inverse probability of censoring weighting (IPCW) is widely used to address dependent censoring in survival analysis. {Beyond estimation of the censoring distribution, its implementation requires specifying how inverse probability weights are incorporated into survival estimators. IPCW does not uniquely determine this construction, and different choices can produce different finite-sample behavior even under the same censoring model.}

{In this work, we examine a class of IPCW estimators for the survival function, distinguishing hazard-based from margin-based formulations and time-varying from fixed weighting schemes. We show how these choices affect structural validity, stability, and finite-sample performance. In particular, the weight-changing hazard (WCH) estimator applies time-varying censoring weights within a product-limit construction with risk-set normalization, ensuring monotonicity and boundedness when the weighted risk-set denominators are positive. Through simulation studies under varying levels of covariate-dependent and administrative censoring, we demonstrate that WCH achieves favorable bias--variance tradeoffs and improved finite-sample stability relative to the alternative IPCW constructions examined, particularly in settings with few observed events or heavy censoring. We further illustrate these differences using data from the Women's Interagency HIV Study (WIHS), in which alternative IPCW constructions produced meaningfully different survival curves.}

Overall, our findings emphasize that the effectiveness of IPCW depends on both the weighting scheme and the estimator formulation. Careful consideration of these choices is therefore important for reliable survival estimation under dependent censoring.

\medskip
\noindent\textbf{Keywords:} informative censoring; inverse probability of censoring weighting; Kaplan--Meier estimator; product-limit estimator; survival analysis; weighted survival estimation

\section{Introduction}
Right-censoring is pervasive in biomedical and epidemiologic studies, and inverse probability of censoring weighting (IPCW) has become a standard tool for addressing bias when censoring depends on observed covariates \citep{robins1992recovery,robins1993information,robins2000correcting}.
The core idea is simple: {under conditionally independent censoring and positivity,} appropriately reweighting observed outcomes using the censoring survival function \(G(t\mid L)\), where \(L\) denotes a vector of observed covariates, can recover target quantities.
{This IPCW perspective is now embedded across a wide range of survival-methods workflows, including covariate-adjusted estimation, time-dependent prediction error assessment, and algorithmic learning with censored outcomes \citep{hernan2000marginal,graf1999assessment,pepe2003statistical,gerds2006consistent,mogensen2012evaluating,prince2025estimation}.
In particular, for estimation of the survival function, censoring weights can be incorporated through either hazard-based product-limit constructions or margin-based weighted averages.}

{The hazard-based approach builds on the classical Kaplan--Meier estimator, which introduced the product-limit construction based on successive risk sets \citep{kaplan1958nonparametric}. Later work developed covariate-adjusted survival curves and weighted product-limit estimators for related adjustment problems \citep{makuch1982adjusted,xie2005adjusted}. In a hazard-based IPCW construction, inverse censoring weights are applied to event counts and at-risk counts to estimate hazard increments, which are then combined through a product limit. Therefore, this approach incorporates censoring adjustment within the risk-set structure of survival estimation.}

{Alternatively, the margin-based approach estimates survival probabilities directly through weighted averages of time-specific outcomes, a construction that arises naturally from a missing-data perspective \citep{robins1995analysis}. Specifically, the survival function \(S(t)=\Pr(T\ge t)\) is the mean of the latent indicator \(Y(t)=\mathbb{I}(T\ge t)\). When \(T\) is subject to censoring by \(C\), this quantity can be estimated using an appropriately inverse-weighted mean of the observable indicator \(\mathbb{I}(\widetilde T\ge t)\), where \(\widetilde T=\min(T,C)\) is the observed time. Weighted-average and product-limit constructions coincide in certain settings, as illustrated by the IPCW representation of the ordinary Kaplan--Meier estimator under independent censoring \citep{satten2001kaplan}. However, this equivalence does not generally extend to covariate-dependent censoring weights, motivating an examination of how different IPCW constructions affect estimator properties even when they use the same estimated censoring distribution.}

{A separate choice concerns the time at which the censoring weight is evaluated. Time-varying weighting uses \(G(t\mid L)\) at each risk-set or evaluation time, whereas fixed weighting assigns each observed failure a single weight based on \(G(\widetilde T\mid L)\). This weighting choice is distinct from whether the resulting estimator is constructed through weighted hazard increments or a weighted marginal mean. Consequently, IPCW survival estimators can differ along two dimensions: the weighting scheme and the estimator construction.}

Recent work comparing Kaplan--Meier--based weighting strategies for regression functionals highlights that similar IPCW approaches can behave quite differently once one accounts for how weights are constructed and used, underscoring the need for explicit estimator definitions and guidance \citep{blanche2013estimating, overgaard2025comparison}.
This paper further argues that estimator choice is not merely a computational detail.
A survival function is not an arbitrary regression curve, and it must be monotone non-increasing and bounded within \([0,1]\).
{Although the classical Kaplan--Meier construction naturally preserves these properties \citep{kaplan1958nonparametric,fleming2013counting,kalbfleisch2002statistical}, their preservation by IPCW estimators depends on both the weighting scheme and the estimator structure.
Beyond these structural properties, different IPCW constructions may also exhibit substantial differences in finite-sample stability as censoring increases and risk sets shrink.}
{Motivated by these considerations, we revisit the IPCW-weighted product-limit estimator
\citep{robins2000correcting, satten2001estimating}, which applies time-varying inverse censoring weights to both the numerator and the
denominator of each weighted hazard increment, and compare it systematically with alternative IPCW constructions. We call it the weight-
changing hazard (WCH) estimator to place it within this class. With finite nonnegative weights and positive weighted risk-set
denominators, WCH is a proper survival curve by construction, and it normalizes each weighted event count by the weighted risk set, which
includes subjects who are censored later. Our goal is to clarify how estimator structure, with the weights held fixed, influences the
validity, stability and finite-sample behavior of IPCW survival estimation. We establish the relationships, support requirements and
probability limits of the estimators formally, and use simulation to show that WCH offers a favorable bias--variance tradeoff and improved
finite-sample stability, especially with few events or heavy dependent censoring, whereas alternative estimators can be unstable or can
target a different quantity even when the censoring model is correctly specified.}

{This work makes three primary contributions. First, we provide a unified taxonomy of IPCW survival estimators, distinguishing margin-based from hazard-based formulations and fixed from time-varying weighting schemes, and establish algebraic relationships among these estimators. Second, we clarify that different positivity assumptions of $G(t \mid L)$ are required for different estimators, and under bounded follow-up, certain estimators may lose unobservable event-time mass. Third, we distinguish these support requirements from curve-shape constraints and finite-sample stability, and evaluate how weighting scheme and construction jointly affect estimator performance, with particular attention to the dynamic risk-set normalization used by WCH.}

The remainder of the paper is organized as follows. In Section 2 (Methods), we introduce notation and formally define a class of IPCW estimators for the survival function, including margin-based and hazard-based constructions with fixed or time-varying weights. We clarify their relationships and discuss key structural properties, such as monotonicity, boundedness, and risk-set-based normalization, that are essential for valid survival estimation. In Section 3 (Simulation), we conduct a comprehensive simulation study to compare the finite-sample performance of these IPCW estimators under a range of censoring regimes, with particular emphasis on settings with heavy or covariate-dependent censoring and limited numbers of observed events.
{In Section 4 (Application), we illustrate the practical implications of estimator choice using publicly available data from the Women's Interagency HIV Study \citep{bacon2005wihs}, demonstrating how different IPCW constructions can lead to meaningfully different survival estimates.}
Finally, Section 5 (Discussion) summarizes the main findings, highlights implications for applied survival analysis, and outlines directions for future work.

\section{Methods}

\subsection{Notation and Setup}

Let \( T \) denote the event time of interest (e.g., death), and \( C \) the censoring time. For each individual \( i = 1, \ldots, n \), we observe the follow-up time \( \tilde{T}_i = \min(T_i, C_i) \), and an event indicator \( \Delta_i = \mathbb{I}(T_i \leq C_i) \). Let \( L_i \in \mathbb{R}^p \) denote a vector of observed baseline covariates. We assume that the observed data, \(O_i = (\tilde{T}_i, \Delta_i, L_i)\), are independent and identically distributed.

Our estimand of interest is the marginal survival function:
\[
S(t) = \Pr(T \ge t),
\] defined on a finite time horizon \(t \in [0, \tau]\). For continuous event times, the distinction between \( >\) and \(\ge\) is immaterial. Throughout the paper we adopt the \( \ge \) notation for consistency. We assume right censoring only and adopt the standard conditional independent censoring assumption \(T \perp C\mid L\), which allows censoring to depend on observed covariates but not on the event time after conditioning on \(L\).

Let
\[
G(t\mid L) = \Pr(C\ge t\mid L)
\] denote the conditional survival function of the censoring distribution. In practice, \(G(t\mid L)\) is unknown and must be estimated; however, in this paper our primary focus is not how \(G\) is estimated, but rather how inverse censoring weights derived from \(G\) are used to construct estimators of \(S(t)\).

\subsection{Inverse Probability of Censoring Weighting}

Inverse probability of censoring weighting (IPCW) addresses dependent censoring by reweighting observed contributions according to the inverse probability of remaining uncensored. Intuitively, individuals who are more likely to be censored early receive larger weights to represent similar individuals who would have remained under observation.
IPCW estimators can be motivated through two related but distinct representations of the survival function, leading naturally to either time-varying or fixed weighting schemes. These constructions are mathematically related but give rise to different finite-sample estimators once embedded into survival analysis procedures.

First, consider estimating the survival function $S(t)=\Pr(T \geq t)$ directly. The survival indicator $\mathbb{I}\{T_i \geq t\}$ is observable whenever the observed time $\tilde{T}_i \geq t$. This motivates a time-specific weighting scheme of the form
\[
w_i(t) = \frac{1}{G(t \mid  L_i)},
\]
which adjusts for the probability of remaining uncensored up to a fixed time $t$. These weights vary with evaluation time \(t\) and therefore adapt dynamically as the effective risk set evolves over follow-up.
Under the standard conditional independence assumptions, the weighting recovers the survival probability
\begin{equation}
\begin{aligned}
    &\mathbb{E}\left[ {w_i(t)}{\mathbb{I}(\tilde{T}_i \geq t)}  \right] \\
    &\quad= \mathbb{E}_{L_i}\left[ \frac{
   \Pr(T_i \geq t \mid  L_i)\Pr(C_i \geq t \mid  L_i)}{G(t\mid L_i)}\right]
    \\
    &\quad=
    \Pr(T_i \geq t),
\end{aligned}
\end{equation}
and similarly recovers the event probability over $[t, t+dt)$
\begin{equation}
\begin{aligned}
    &\mathbb{E}\left[{w_i(t)} \Delta_i{ \mathbb{I}(t \leq \tilde{T}_i < t+dt)}  \right] \\
    &\quad= \mathbb{E}_{L_i, T_i}\left[ \frac{G(T_i \mid  L_i)
    \mathbb{I}(t \leq T_i < t+dt)}{G(t\mid L_i)}\right] \\
    &\quad=
    \Pr(t \leq T_i < t+dt),
\end{aligned}
\end{equation}
where the second identity holds when $dt \rightarrow 0$, assuming $G(t \mid L)$ is continuous in $t$. Analogous results hold in the discrete-time setting.

Alternatively, one may express the survival function as $S(t)=1-\Pr(T_i < t)$ and instead focus on indicators $\mathbb{I}\{T_i < t\}$, which are observable when failure occurs before $t$, i.e., $\Delta_i=1$ and $\tilde{T}_i < t$. This leads to weights of the form
\[
w_i = \frac{1}{{G}(\tilde{T}_i \mid  L_i)},
\]
assigning each individual a single weight evaluated at their observed follow-up time. Unlike the previous construction, these weights are evaluated once at each individual’s observed follow-up time and remain fixed across all subsequent estimation steps. As a result, fixed-weight IPCW estimators do not directly adapt to changes in the risk set over time.
In particular,
\begin{align}
\nonumber
    \mathbb{E}
    \left[
    { w_i \Delta_i \mathbb{I}\{\tilde{T}_i < t\}}
    \right]  & =
    \mathbb{E}_{T_i, L_i}
    \left[
    \mathbb{E}_{C_i}
    \left[
    \mathbb{I}\{T_i < t \}
    \frac{ \mathbb{I}\{T_i \leq C_i\}}{{G}( T_i \mid  L_i)}
    \mid  L_i, T_i
    \right]
    \right]
    \\
    \label{eq::sym1}
    & = \Pr(T_i < t)
\end{align}
and similarly
\begin{align}
\nonumber
        \mathbb{E}
        \left[
        w_i \Delta_i \mathbb{I}\{t\leq \tilde{T}_i < t+dt\}
        \right]
        = & \mathbb{E}_{T_i, L_i}\left[
        \frac{G(T_i \mid  L_i)}{{G}(T_i \mid  L_i)}
        \mathbb{I}\{ t \leq T_i < t+dt\}
        \right] \\
        = &  \Pr(t \leq T_i < t+dt),
\end{align}
Moreover, this weighting scheme exhibits a useful symmetry in the sense that
\begin{align}
\nonumber
    \mathbb{E}
    \left[
    { w_i\Delta_i\mathbb{I}\{\tilde{T}_i \geq t\}}
    \right]
     = &
    \mathbb{E}_{T_i, L_i}
    \left[
    \mathbb{E}_{C_i}
    \left[
    \frac{ \mathbb{I}\{t \leq T_i \leq C_i\}}{{G}( T_i \mid  L_i)}
    \mid  L_i, T_i
    \right]
    \right] \\
    \label{eq::sym2}
    =  & \Pr(T_i\geq t)
\end{align}
so that both representations, either based on $1-\mathbb{I}\{\tilde{T}_i < t\}$ or $\mathbb{I}\{\tilde{T}_i \geq t\}$, yield valid moment conditions for the underlying survival function.

Although both weights arise naturally from missing-data arguments, they lead to distinct estimators of the survival function once embedded into survival analysis procedures. In particular, IPCW does not uniquely define an estimator of \(S(t)\), and additional choices regarding estimator structure are required. We make these choices explicit below.

\subsection{IPCW Estimators of the Survival Function}
We consider a class of IPCW estimators that vary along two primary dimensions:
1) \textbf{Weighting scheme}: whether weights are time-varying or fixed, and
2) \textbf{Estimator construction}: whether estimation proceeds via hazard-based (product-limit) or margin-based (pointwise) construction. These distinctions define four primary estimator classes: weight-changing hazard (WCH), weight-changing margin (WCM), weight-fixed hazard (WFH) and weight-fixed margin (WFM). In addition, we consider alternative representations (WFH-2 and WFM-2) derived from the complementary representation of the survival function.

\begin{itemize}
    \item[1.] \textbf{Weight-Changing Hazard Estimator (WCH)}: The weight-changing hazard estimator applies time-specific IPC weights within a product-limit (hazard-based) framework:
    \begin{equation} \label{eq::hazard-WCH}
    \hat \lambda_{\operatorname{WCH}}(s) = \frac{\sum_{i=1}^n w_i(s)\Delta_i\mathbb{I} (\tilde{T}_i =s)  }{\sum_{i=1}^n  w_i(s)\mathbb{I} (\tilde{T}_i \ge s) },
    \end{equation}
    \[
    \hat S_{\operatorname{WCH}}(t) = \prod_{s \leq t} \left( 1 - \hat \lambda_{\operatorname{WCH}}(s) \right).
    \]
This estimator extends the product-limit structure of the Kaplan–Meier estimator by incorporating time-varying inverse censoring weights within weighted hazard increments.
    \item[2.] \textbf{Weight-Changing Margin Estimator (WCM):}
    A margin-based estimator using time-varying weights is
    \[
    \hat S_{\operatorname{WCM}}(t) = \frac{1}{n}\sum_{i=1}^n w_i(t)
    {\mathbb{I}(\tilde{T}_i\geq t) },
    \]
    which corresponds to a Horvitz–Thompson estimator applied pointwise in time. Because estimation proceeds independently at each evaluation time point, this estimator does not inherently enforce monotonicity or boundedness of the resulting survival curve.
    \item[3.] \textbf{Weight-Fixed Hazard Estimator (WFH):} Using weights evaluated at observed follow-up times yields fixed-weight IPCW estimators, which can be interpreted as empirical estimators of the event distribution based solely on observed failures.
    For example, using the fixed weights $w_i = 1/G(\tilde{T}_i \mid  L_i)$, a product-limit–type estimator can be defined as
    \begin{equation} \label{eq::hazard-WFH}
    \hat \lambda_{\text{WFH}}(s) = \frac{\sum_{i=1}^n w_i \Delta_i  \mathbb{I}(\tilde{T}_i = s)}{n-\sum_{i=1}^n w_i \Delta_i  \mathbb{I}(\tilde{T}_i < s)},
    \end{equation}
    \[
    \hat{S}_{\text{WFH}}(t) = \prod_{s \leq t} \left(1 - \hat \lambda_{\text{WFH}}(s) \right).
    \]
    \item[4.] \textbf{Weight-Fixed Margin Estimator (WFM): }
    Based on the formula in (\ref{eq::sym1}), a margin-based estimator using fixed weights can naturally be defined as
    \[
    \hat S_{\text{WFM}}(t) = 1- \frac{1}{n}\sum_{i=1}^n w_i\Delta_i
    {\mathbb{I}(\tilde{T}_i< t) }.
    \]
    \item[5.] \textbf{Weight-Fixed Hazard Estimator-2 (WFH-2):}
    To better understand the role of normalization in fixed-weight IPCW estimation, we additionally consider alternative normalized formulations derived from the representation \( S(t) = \Pr(T \ge t)\):
     \begin{equation} \label{eq::hazard-WFH2}
    \hat \lambda_{\text{WFH-2}}(s) = \frac{\sum_{i=1}^n w_i \Delta_i \mathbb{I}(\tilde{T}_i = s)}{\sum_{i=1}^n w_i \Delta_i \mathbb{I}(\tilde{T}_i \geq s)},
    \end{equation}
    \[
    \hat{S}_{\text{WFH-2}}(t) = \prod_{s \leq t} \left(1 - \hat \lambda_{\text{WFH-2}}(s) \right).
    \]
    Unlike WFH, which builds the denominator of the hazard through estimating   $1-\Pr(T_i < t)$, this formulation directly estimates $\Pr(T_i \geq t)$.
    \item[6.] \textbf{Weight-Fixed Margin Estimator-2 (WFM-2): }
    Correspondingly, an alternative margin-based estimator is obtained from the complementary representation of the survival function:
    \[
    \hat S_{\text{WFM-2}}(t) = \frac{1}{n}\sum_{i=1}^n w_i\Delta_i
    {\mathbb{I}(\tilde{T}_i\geq t) }.
    \]
    When their respective positivity requirements hold, WFM and WFM-2 (and similarly WFH and WFH-2) target the same population quantity, but they rely on different empirical representations and are generally not identical in finite samples.
\end{itemize}

\subsection{Structural Properties and Implications}

\sh{In this subsection, we examine the relationships among the IPCW estimators, along with their structural properties for identification, monotonicity, boundedness, and finite-sample stability.}

\paragraph{Relationships among estimators.}

\sh{Several constructions are closely related. First, the two fixed-weight margin estimators are connected through
\[
\hat S_{\mathrm{WFM}}(t)-\hat S_{\mathrm{WFM-2}}(t)
=1-\frac{1}{n}\sum_{i=1}^n w_i\Delta_i
=1-\bar w,
\]
where
\[
\bar w=\frac{1}{n}\sum_{i=1}^n w_i\Delta_i.
\]
In other words, the two curves differ by a constant vertical shift determined by the total weighted failure mass, regardless of the evaluation time.
}

\sh{Second, the corresponding fixed-weight hazard constructions can also be re-expressed through weighted averages. For the WFH and WFM estimators, let \(s_t\) and \(s_{t+1}\) denote consecutive distinct event times with \(s_t<t<s_{t+1}\). Provided the denominators are nonzero, the product defining WFH telescopes as
\begin{align*}
\hat{S}_{\text{WFH}}(t)
&= \prod_{s \leq t}
\frac{n-\sum_{i=1}^n w_i\Delta_i\mathbb{I}\{\tilde{T}_i\leq s\}}
{n-\sum_{i=1}^n w_i\Delta_i\mathbb{I}\{\tilde{T}_i<s\}} \\
&= \frac{n-\sum_{i=1}^n w_i\Delta_i\mathbb{I}\{\tilde{T}_i\leq s_t\}}{n} \\
&= 1-\frac{1}{n}\sum_{i=1}^n w_i\Delta_i\mathbb{I}\{\tilde{T}_i<t\}
= \hat{S}_{\text{WFM}}(t),
\end{align*}
where the product in the first line is over distinct event times $s$. At the event times,
\begin{align*}
\hat{S}_{\text{WFH}}(s_t)
&= 1-\frac{1}{n}\sum_{i=1}^n w_i\Delta_i\mathbb{I}\{\tilde{T}_i\leq s_t\} \\
&= 1-\frac{1}{n}\sum_{i=1}^n w_i\Delta_i\mathbb{I}\{\tilde{T}_i<s_{t+1}\}
= \hat{S}_{\text{WFM}}(s_{t+1}).
\end{align*}
Thus, the estimators agree between event times and differ at the jumps only through their endpoint convention. The product-limit expression therefore does not guarantee boundedness of WFH.}

\sh{Similarly, the relationship between WFH-2 and WFM-2 can be established through weight normalization. When \(\bar w>0\), normalizing the fixed weights through
\[
\tilde w_i=\frac{w_i}{\bar w}
\]
makes the empirical mean of the weighted failure contributions equal to one:
\[
\frac{1}{n}\sum_{i=1}^n\tilde w_i\Delta_i=1.
\]
Define the normalized WFM-2 estimator as
\[
\hat S_{\mathrm{WFM-2-n}}(t)
=\frac{1}{n}\sum_{i=1}^n\tilde w_i\Delta_i\mathbb{I}(\tilde T_i\ge t)
=\frac{\sum_{i=1}^n w_i\Delta_i\mathbb{I}(\tilde T_i\ge t)}{\sum_{i=1}^n w_i\Delta_i}.
\]
For \(t\) between consecutive event times, with \(s_t\) denoting the last event time before \(t\), the product defining WFH-2 telescopes as
\begin{align*}
\hat S_{\mathrm{WFH-2}}(t)
&=\prod_{s\leq t}
\frac{\sum_{i=1}^n w_i\Delta_i\mathbb{I}(\tilde T_i>s)}
{\sum_{i=1}^n w_i\Delta_i\mathbb{I}(\tilde T_i\geq s)} \\
&=\frac{\sum_{i=1}^n w_i\Delta_i\mathbb{I}(\tilde T_i>s_t)}
{\sum_{i=1}^n w_i\Delta_i} \\
&=\frac{1}{n}\sum_{i=1}^n
\frac{w_i\Delta_i\mathbb{I}(\tilde T_i\geq t)}{\bar w}
=\hat S_{\mathrm{WFM-2-n}}(t),
\end{align*}
where the product is over distinct event times and requires nonzero denominators. At event times, the two expressions differ only through their endpoint convention, as with WFH and WFM:
\[
\hat S_{\mathrm{WFH-2}}(s_t)
=\hat S_{\mathrm{WFM-2-n}}(s_{t+1}).
\]
Thus, with aligned event-time conventions, WFH-2 is a H\'ajek-type normalization of WFM-2.}

\paragraph{Positivity assumptions and follow-up support.}
\sh{Identification of \(S(t)\) through IPCW requires conditional independent censoring and correct specification, along with the appropriate positivity assumptions for \(G(t\mid L)\), but the required positivity conditions differ across estimators.
For a given evaluation time \(t\), WFM-2 and WFH-2 require positivity across the event-time support, whereas WFH, WFM, WCM, and WCH require positivity only at or before the evaluation time \(t\).
Thus, bounded follow-up due to administrative censoring can violate the positivity requirements of WFM-2 and WFH-2 even when those of the other constructions remain satisfied.}

\sh{To make the positivity requirements explicit, assume conditional independent censoring and consider weights based on the true censoring distribution. At an observed failure time \(\tilde T_i=T_i\), if \(G(T_i\mid L_i)>0\) almost surely, then
\[
\mathbb{E}[w_i\Delta_i\mid T_i,L_i]
=\frac{\Pr(C_i\ge T_i\mid T_i,L_i)}{G(T_i\mid L_i)}
=\frac{G(T_i\mid L_i)}{G(T_i\mid L_i)}
=1.
\]
Taking expectations then gives
\[
\mathbb{E}[w_i\Delta_i]=1,
\qquad
\mathbb{E}\!\left[w_i\Delta_i\mathbb{I}(\tilde T_i\ge t)\right]
=\mathbb{E}\!\left[\mathbb{I}(T_i\ge t)\right]
=S(t).
\]
Therefore, positivity across the event-time support validates the WFM-2 moment representation. Under regularity conditions, WFH-2 has the same probability limit because its normalizing factor \(\bar w\) converges to one. }

\sh{However, this cancellation fails when failures at some possible event times have zero probability of being observed. To illustrate, suppose the censoring distribution has a finite endpoint \(\tau_{\max}\), with \(G(u\mid L)>0\) for \(u<\tau_{\max}\) and \(G(u\mid L)=0\) for \(u>\tau_{\max}\), and assume \(\Pr(T=\tau_{\max})=0\). For an individual whose event occurs beyond \(\tau_{\max}\), \(\Delta_i=0\) with probability one, so the weighted failure contribution is always zero. The conditional expectation above therefore becomes
\[
\mathbb{E}[w_i\Delta_i\mid T_i,L_i]
=\mathbb{I}(T_i<\tau_{\max}),
\]
rather than one. Consequently,
\[
\mathbb{E}[w_i\Delta_i]=\Pr(T_i<\tau_{\max}),
\]
and, for \(t<\tau_{\max}\),
\[
\mathbb{E}\!\left[w_i\Delta_i\mathbb{I}(\tilde T_i\ge t)\right]
=\Pr(t\le T_i<\tau_{\max}).
\]
Therefore, the proposed WFM-2 is biased, even with a known censoring distribution and increasing sample size.
Additionally, normalizing WFM-2 does not recover the missing mass, and WFH-2 converges to
\[
\frac{\Pr(t\le T<\tau_{\max})}{\Pr(T<\tau_{\max})}
=\Pr(T\ge t\mid T<\tau_{\max}).
\]
provided that \(\Pr(T<\tau_{\max})>0\).}

\sh{In contrast, WFM requires positivity only at failure times before \(t\). For \(t<\tau_{\max}\),
\[
\mathbb{E}\!\left[w_i\Delta_i\mathbb{I}(\tilde T_i<t)\right]
=\Pr(T_i<t),
\]
and hence
\[
1-\mathbb{E}\!\left[w_i\Delta_i\mathbb{I}(\tilde T_i<t)\right]=S(t).
\]
WFH shares this support requirement through its algebraic relationship with WFM, with endpoint conventions aligned and the product defined.}

\sh{For WCM or WCH, positivity is required at the evaluation time itself. Since \(\tilde T_i=\min(T_i,C_i)\), conditional independent censoring gives
\[
\mathbb{E}\!\left[\frac{\mathbb{I}(\tilde T_i\ge t)}{G(t\mid L_i)}\middle|L_i\right]
=\frac{\Pr(T_i\ge t,C_i\ge t\mid L_i)}{G(t\mid L_i)}
=\Pr(T_i\ge t\mid L_i),
\]
provided \(G(t\mid L_i)>0\). Averaging over \(L_i\) then yields \(S(t)\). Thus, under the required assumptions, WFM, WFH, WCM, and WCH can target \(S(t)\) within bounded follow-up even when event-time mass remains beyond the follow-up endpoint.}

\sh{
Finally, zero observation probabilities should be distinguished from small but positive probabilities. The former can invalidate the moment representation, whereas the latter can produce large weights and finite-sample instability even when the population identity remains valid.}

\paragraph{Monotonicity and boundedness.}

\sh{Identification and the shape constraints of a survival curve are distinct properties. With finite nonnegative weights and positive weighted risk-set denominators, WCH and WFH-2 have hazard increments in \([0,1]\), as shown in equations~(\ref{eq::hazard-WCH}) and~(\ref{eq::hazard-WFH2}), yielding non-increasing curves bounded in \([0,1]\). However, these shape guarantees do not ensure that WFH-2 targets the marginal survival function under bounded follow-up. Although WFH also has a product-limit form, its hazard increments in equation~(\ref{eq::hazard-WFH}) need not lie in \([0,1]\), so boundedness is not guaranteed. Nevertheless, wherever its product is defined, WFH is non-increasing through its algebraic relationship with WFM.}

\sh{The structural limitations of the margin-based estimators also differ. With fixed nonnegative weights, both WFM and WFM-2 are non-increasing. However, WFM can fall below zero if the cumulative weighted failure mass exceeds one, whereas WFM-2 can exceed one if its weighted tail mass exceeds one. By contrast, time-varying weighted means WCM need not be monotone or bounded by one. Monotonicity and boundedness depend jointly on the specific weighting and normalization, not simply on whether the estimator has a margin-based or product-limit form.}

\paragraph{Finite-sample stability.}
\sh{For WFM-2, even when the required support conditions hold, finite samples may contain few or no observed failures in the event-time tail. Missing these contributions can leave \(\bar w<1\), placing WFM-2 below WFM by \(1-\bar w\). By contrast, when such failures are observed, their small observation probabilities lead to large inverse weights, which can dominate the weighted tail sum and cause substantial finite-sample variability. WFH-2 normalizes WFM-2 by the total weighted failure mass, potentially improving stability, but this normalization neither guarantees better finite-sample performance nor recovers unobservable tail mass. WFM and its hazard-based counterpart WFH do not depend on failures after the evaluation time, although large weights on earlier failures can still cause instability. WCH instead evaluates censoring weights at each event time and normalizes weighted event counts by the corresponding weighted at-risk counts. This dynamic risk-set normalization preserves monotonicity and boundedness when the denominators are positive and may improve finite-sample stability. We examine the relative performance of these estimators in the following simulation study.}

\sh{The following simulation study evaluates these differences across several censoring scenarios. We distinguish settings in which the estimators target the same survival function from settings in which bounded support changes the targets of WFM-2 and WFH-2. This distinction allows us to compare curve shape, bias, and variability while separating finite-sample behavior from discrepancies caused by unobservable event-time mass beyond the follow-up horizon.}

\section{Simulation}

\sh{The simulation study used two matched settings that differed only by the presence of a fixed administrative censoring time. This design separates finite-sample behavior under full support from the estimand change induced by a bounded observation horizon.}

\subsection{Data Generation}
\sh{
Two baseline covariates, $L_1$ and $L_2$, were independently drawn from
a uniform distribution on $[-1,1]$. We used the Weibull shape--scale
parameterization, under which a distribution with shape $\kappa$ and
scale $\lambda$ has survival function
$\exp\{-(t/\lambda)^\kappa\}$. Event times were generated according to
\[
T\mid L\sim\operatorname{Weibull}\{2,\lambda_T(L)\},
\qquad
\lambda_T(L)=\exp\{-2-L_2-2|L_1|\}^{-1/2},
\]
and the covariate-dependent informative censoring times were generated according to
\[
\begin{gathered}
C^R\mid L\sim\operatorname{Weibull}\{1,\lambda_C(L)\},
\\
\lambda_C(L)
=
\exp\{\alpha_C-1.2L_2-0.8|L_1|\}^{-1}.
\end{gathered}
\]
We considered $\alpha_C=-0.47,-1.09,$ and $-1.61$, corresponding to
higher, moderate, and lower levels of informative censoring, respectively.
}

\sh{In Setting I, $C=C^R$, yielding approximately 27\%, 45\%, and 60\% observed events under higher, moderate, and lower censoring, respectively. The conditional Weibull censoring distribution has positive survival at every finite time, so all the IPCW estimators satisfy the required positivity assumptions. This setting evaluates finite-sample differences when the estimators share the same population target.}

\sh{In Setting II, an administrative censoring time of 6 was additionally imposed on every individual, so that $C=\min(C^R,6)$; the corresponding observed-event proportions were approximately 23\%, 37\%, and 48\%. The event-time and informative censoring distributions were otherwise unchanged. All evaluation times, $t=1,3,5$, preceded this endpoint. Thus, positivity through each evaluation time was preserved, while failures beyond time 6 could not be observed. This setting examines how the additional support requirement of the failure-only tail representations affects their population targets.}

\subsection{Estimators Compared \& Evaluation Metrics}

We compared all IPCW estimators defined in Section 2. The naive Kaplan--Meier estimator (KM) was also included as the baseline comparator.
In all cases, IPC weights were constructed using the true censoring survival function \(G(t\mid L)\), eliminating bias due to weight estimation.
\sh{Each simulation setting was repeated $M=1000$ times for sample sizes $n=100,500,5000$. Performance was evaluated at $t=1,3,5$.}

At each time point, performance was evaluated using the standard metrics, including bias (\textbf{bias}), mean absolute error (\textbf{MAE}), mean absolute relative error (\textbf{MARE}), standard deviation (\textbf{sd}), and root mean squared error (\textbf{rmse}).
Here, {mean absolute error} denotes the average absolute difference between the estimated and reference survival probabilities across simulation replications, and {mean absolute relative error} is that quantity divided by the reference survival probability.
In addition, we reported the proportions of estimates falling within the valid range $[0,1]$ (\textbf{Range}), and the proportion exhibiting monotonicity across the three time points $t=1,3,5$ within each simulation (\textbf{Monotonicity}).

\subsection{Results}

\begin{figure}[htbp]
    \centering
    \includegraphics[width=0.97\textwidth,height=0.78\textheight,keepaspectratio]{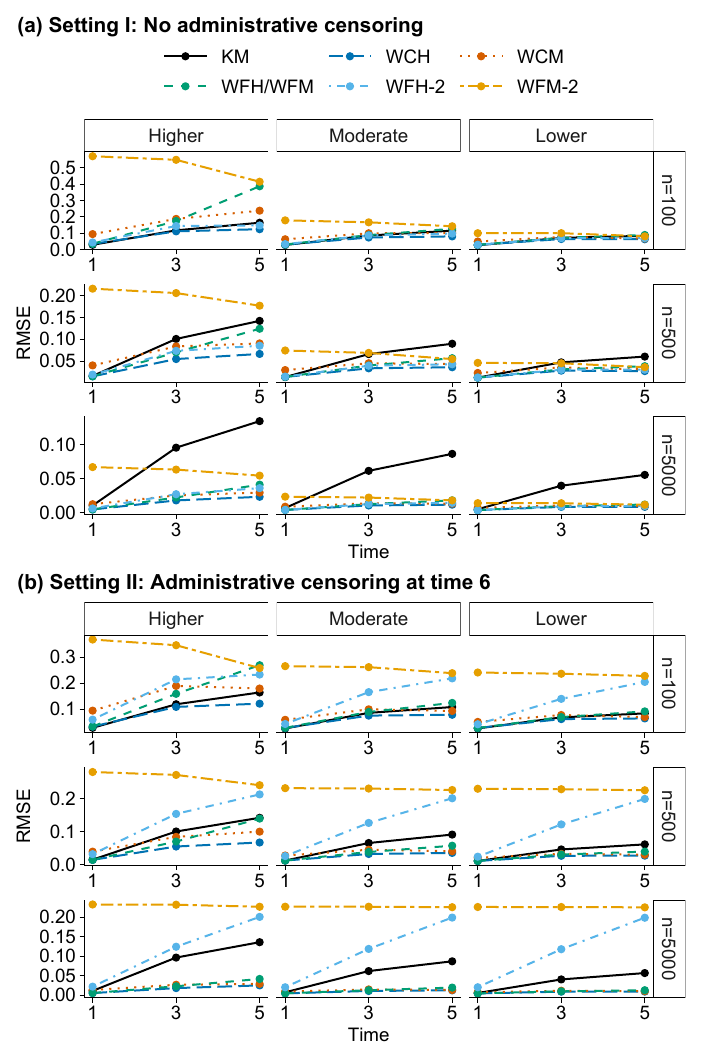}

    \caption{\sh{RMSE across evaluation times, censoring levels, and sample sizes $n=100,500,5000$ for Setting I (top) and Setting II (bottom).}}
    \label{fig:simulation-rmse}
\end{figure}

\begin{figure}[htbp]
    \centering
    {\small\bfseries
    \sh{(a) Setting I: No administrative censoring}\par}
    \smallskip
    \includegraphics[width=\textwidth,height=0.32\textheight,keepaspectratio]
        {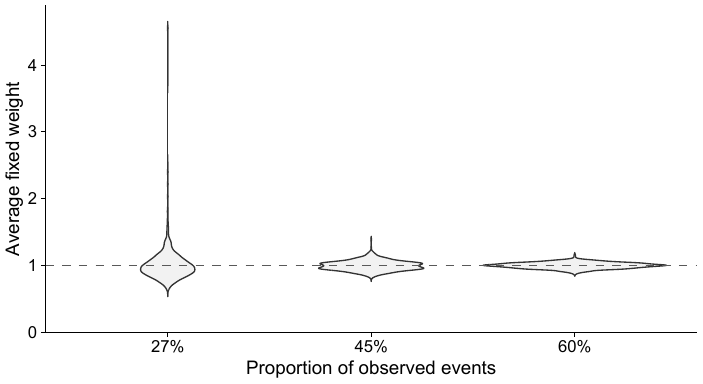}
    \par\medskip

    {\small\bfseries
    \sh{(b) Setting II: Administrative censoring at time 6}\par}
    \smallskip
    \includegraphics[width=\textwidth,height=0.32\textheight,keepaspectratio]
        {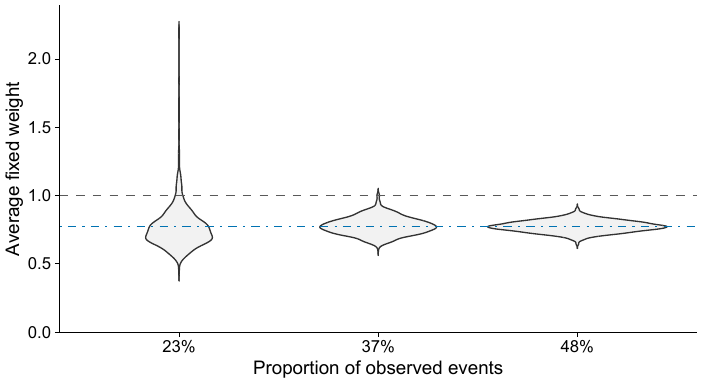}

    \caption{\sh{Distributions of the averaged fixed weights, $\bar w$, across simulations with $n=500$ for Setting I (top) and Setting II (bottom). The dashed horizontal line denotes one; the dot-dashed line in Setting II denotes $1-S(6)=0.7755$}}
    \label{fig:simulation-fixed-weights}
\end{figure}

\begin{figure}[htbp]
    \centering
    {\small\bfseries
    \sh{(a) Setting I: No administrative censoring}\par}
    \smallskip
    \includegraphics[width=\textwidth,height=0.32\textheight,keepaspectratio]
        {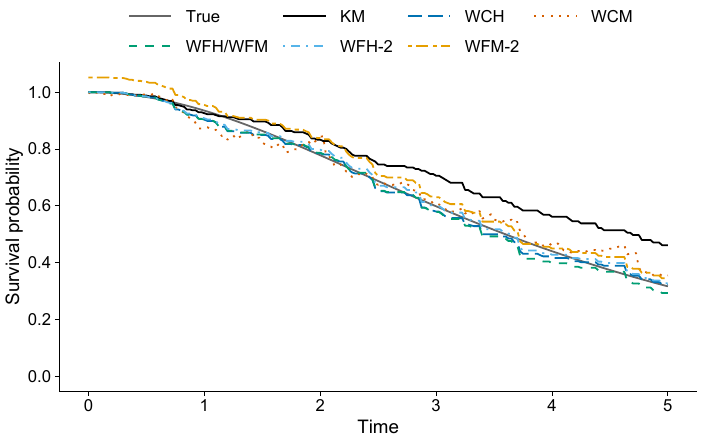}
    \par\medskip

    {\small\bfseries
    \sh{(b) Setting II: Administrative censoring at time 6}\par}
    \smallskip
    \includegraphics[width=\textwidth,height=0.32\textheight,keepaspectratio]
        {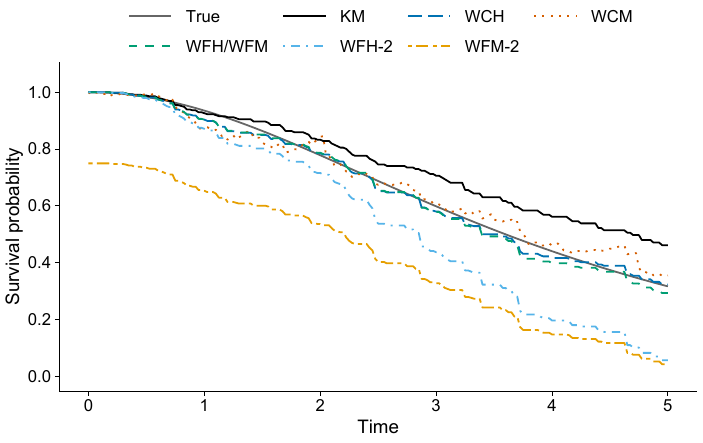}

    \caption{\sh{Estimated survival curves from representative samples with $n=500$ under higher censoring for Setting I (top) and Setting II (bottom). The vertical axis includes estimates above one}}
    \label{fig:simulation-one-time}
\end{figure}

\sh{Tables~\ref{tab:weibull-full} and~\ref{tab:weibull-admin} summarize the simulation results for the two settings. Figure~\ref{fig:simulation-rmse} displays RMSE across censoring levels, sample sizes, and evaluation times; Figure~\ref{fig:simulation-fixed-weights} presents the distributions of the averaged fixed weights $\bar{w}$ across simulations; and Figure~\ref{fig:simulation-one-time} shows the complete estimated survival curves from representative simulated samples.}

\sh{Overall, WCH provided the most favorable performance in terms of mean absolute error and RMSE, with its advantage most apparent at later evaluation times, under heavier censoring, and in smaller samples. In Setting I, under higher censoring with $n=100$ at $t=5$, WCH had a bias of 0.0216 and an RMSE of 0.1235, compared with the larger RMSE values of 0.2368 for WCM, 0.3861 for WFH/WFM, 0.1465 for WFH-2, and 0.4138 for WFM-2. These advantages persisted in Setting II because imposing the administrative endpoint did not change the target of WCH. The WCH estimator maintained stable performance across the scenarios considered, including the more challenging settings.}

\sh{WCM and WFH/WFM generally had greater variability than WCH, particularly under heavier censoring and at later follow-up times. WCM uses time-specific inverse weights without any normalization, whereas WFH/WFM accumulates weighted failures before the evaluation time. Both can therefore be sensitive to a small number of large weighted contributions. For example, in Setting I under moderate censoring with $n=500$ at $t=5$, the empirical standard deviations were 0.0428 for WCM and 0.0563 for WFH/WFM, compared with 0.0356 for WCH. WCH instead normalizes the weighted event count by the corresponding weighted risk set at each event time and therefore has more stable performance. The same relative pattern remained in Setting II.}

\sh{The behavior of WFH-2 and WFM-2 depended strongly on the available event-time support. In Setting I, both estimators have the correct population target, but their finite-sample behavior differed because WFH-2 normalizes the fixed weights by the realized sample-average failure weight $\bar{w}$. Under higher censoring with $n=500$ at $t=5$, WFM-2 had an RMSE of 0.1769, whereas WFH-2 had an RMSE of 0.0849. This normalization substantially reduced variability, although it may introduce finite-sample ratio bias and did not guarantee uniformly better performance. As for Setting II, WFH-2 and WFM-2 no longer target the marginal survival function for the reasons described in Section~2.4. Consequently, at $t=5$, WFM-2 converges to approximately
\[
S(5)-S(6)\approx0.0920,
\]
whereas WFH-2 converges to
\[
\frac{S(5)-S(6)}{1-S(6)}\approx0.1186,
\]
rather than to $S(5)\approx0.3166$. At $n=5000$, the WFM-2 means were 0.0917, 0.0918, and 0.0920 under higher, moderate, and lower censoring, respectively. The corresponding WFH-2 means were 0.1182, 0.1184, and 0.1187, illustrating this phenomenon.}

\sh{The distributions of the averaged fixed weights in Figure~\ref{fig:simulation-fixed-weights} clarify the difference between finite-sample instability and loss of support. In Setting I, the population mean of $\bar w$ is approximately one, but finite samples may miss late failures with small observation probabilities. Consequently, the realized mean can be smaller than one, as seen in the plots for $n=500$, potentially shifting WFM-2 downward. In Setting II, by contrast, $\bar w$ was centered near
\[
P(T\leq6)=1-S(6)=0.7755,
\]
because event-time mass beyond the administrative endpoint was systematically unobservable. Increasing the sample size reduced the variability of WFH-2 and WFM-2 but did not eliminate their discrepancy from $S(t)$.}

\sh{The estimators also differed in structural validity. WCH remained monotone and bounded in $[0,1]$ in every replication. WFH-2 was also monotone and bounded, but its behavior in Setting II demonstrates that a structurally valid curve need not be consistent for the intended survival function. WCM could be nonmonotone because it estimates survival separately at each evaluation time. WFH/WFM could fall below zero, whereas WFM-2 could exceed one. Under higher censoring with $n=100$ in Setting I, WCM was nonmonotone across $t=1,3,5$ in 9.2\% of replications. Across all Setting I scenarios, the lowest range-validity proportions were 78.3\% for WCM, 91.4\% for WFH/WFM, and 67.7\% for WFM-2. KM remained structurally valid but was positively biased because it ignored covariate-dependent censoring.}

\sh{Figure~\ref{fig:simulation-one-time} illustrates these behaviors in one representative replicate with $n=500$ under higher censoring. In both settings, KM remained above the true curve, whereas WCH tracked the truth while remaining monotone and bounded. WCM exhibited clear nonmonotonicity. In Setting II, WFH-2 and WFM-2 were shifted downward toward their support-restricted targets. These individual curves illustrate the mechanisms underlying the simulation results.}

\sh{Overall, WCH was comparatively stable across both settings and performed particularly well under heavier censoring and at later follow-up times. Setting I demonstrates its finite-sample advantage when all IPCW estimators share the same population target. Setting II further demonstrates that the advantages of WCH persist under administrative censoring because it requires positivity only through the evaluation time, rather than over the subsequent event-time support. The dynamic risk-set normalization and weaker support requirement of WCH therefore help explain its better performance in the scenarios examined.}

\newgeometry{margin=10mm}
\begin{landscape}
\pagestyle{empty}
\begingroup
\setlength{\intextsep}{0pt}
\begin{table}[H]
\centering
\caption{\sh{Simulation results for Setting I at $n=100,500,5000$. MARE is mean absolute relative error; range is the percentage in $[0,1]$; Mono. assesses the three evaluation times.}}\label{tab:weibull-full}
\fontsize{8}{8.75}\selectfont
\setlength{\tabcolsep}{3.2pt}
\renewcommand{\arraystretch}{0.94}
\begin{adjustbox}{max width=\linewidth}
\begin{tabular}{c|c|c|cccccc|cccccc|cccccc|c}
\toprule
\multirow{2}{*}{Scheme} & \multirow{2}{*}{$n$} & \multirow{2}{*}{Method} & \multicolumn{6}{c|}{$t=1,\ S(t)=0.9351$} & \multicolumn{6}{c|}{$t=3,\ S(t)=0.5985$} & \multicolumn{6}{c|}{$t=5,\ S(t)=0.3166$} & \multirow{2}{*}{Mono.} \\
\cline{4-21}
 & & & bias & MAE & sd & rmse & MARE & range & bias & MAE & sd & rmse & MARE & range & bias & MAE & sd & rmse & MARE & range & \\
\hline
\multirow{18}{*}{\shortstack{Higher censoring\\(27\% events)}} & \multirow{6}{*}{$n=100$} & KM & 0.0090 & 0.0231 & 0.0266 & 0.0280 & 2.5\% & 100.0\% & 0.0934 & 0.0990 & 0.0696 & 0.1164 & 16.5\% & 100.0\% & 0.1373 & 0.1422 & 0.0880 & 0.1631 & 44.9\% & 100.0\% & 100.0\% \\
 &  & WCH & -0.0011 & 0.0266 & 0.0333 & 0.0333 & 2.8\% & 100.0\% & 0.0056 & 0.0868 & 0.1103 & 0.1104 & 14.5\% & 100.0\% & 0.0216 & 0.0995 & 0.1216 & 0.1235 & 31.4\% & 100.0\% & 100.0\% \\
 &  & WCM & -0.0052 & 0.0745 & 0.0934 & 0.0935 & 8.0\% & 78.3\% & -0.0016 & 0.1365 & 0.1865 & 0.1865 & 22.8\% & 97.3\% & 0.0041 & 0.1281 & 0.2369 & 0.2368 & 40.5\% & 98.6\% & 90.8\% \\
 &  & WFM & -0.0015 & 0.0271 & 0.0343 & 0.0343 & 2.9\% & 100.0\% & -0.0069 & 0.1173 & 0.1735 & 0.1735 & 19.6\% & 99.2\% & -0.0152 & 0.2016 & 0.3860 & 0.3861 & 63.7\% & 91.4\% & 100.0\% \\
 &  & WFH-2 & -0.0081 & 0.0326 & 0.0417 & 0.0425 & 3.5\% & 100.0\% & -0.0303 & 0.1137 & 0.1400 & 0.1431 & 19.0\% & 100.0\% & -0.0248 & 0.1169 & 0.1445 & 0.1465 & 36.9\% & 100.0\% & 100.0\% \\
 &  & WFM-2 & 0.0102 & 0.2815 & 0.5701 & 0.5699 & 30.1\% & 69.6\% & 0.0047 & 0.2522 & 0.5482 & 0.5479 & 42.1\% & 93.3\% & -0.0036 & 0.1720 & 0.4140 & 0.4138 & 54.3\% & 98.2\% & 100.0\% \\
\cline{2-22}
 & \multirow{6}{*}{$n=500$} & KM & 0.0102 & 0.0130 & 0.0117 & 0.0155 & 1.4\% & 100.0\% & 0.0957 & 0.0957 & 0.0308 & 0.1005 & 16.0\% & 100.0\% & 0.1358 & 0.1359 & 0.0414 & 0.1420 & 42.9\% & 100.0\% & 100.0\% \\
 &  & WCH & 0.0001 & 0.0115 & 0.0146 & 0.0146 & 1.2\% & 100.0\% & 0.0019 & 0.0432 & 0.0543 & 0.0543 & 7.2\% & 100.0\% & 0.0089 & 0.0529 & 0.0655 & 0.0661 & 16.7\% & 100.0\% & 100.0\% \\
 &  & WCM & -0.0008 & 0.0322 & 0.0398 & 0.0398 & 3.4\% & 94.9\% & 0.0003 & 0.0645 & 0.0834 & 0.0833 & 10.8\% & 99.9\% & 0.0018 & 0.0613 & 0.0903 & 0.0903 & 19.4\% & 99.9\% & 98.7\% \\
 &  & WFM & 0.0001 & 0.0117 & 0.0148 & 0.0148 & 1.3\% & 100.0\% & -0.0005 & 0.0536 & 0.0698 & 0.0697 & 9.0\% & 100.0\% & 0.0017 & 0.0934 & 0.1240 & 0.1239 & 29.5\% & 98.5\% & 100.0\% \\
 &  & WFH-2 & -0.0018 & 0.0145 & 0.0186 & 0.0186 & 1.6\% & 100.0\% & -0.0082 & 0.0575 & 0.0727 & 0.0732 & 9.6\% & 100.0\% & -0.0050 & 0.0650 & 0.0848 & 0.0849 & 20.5\% & 100.0\% & 100.0\% \\
 &  & WFM-2 & 0.0011 & 0.1369 & 0.2160 & 0.2159 & 14.6\% & 72.0\% & 0.0006 & 0.1247 & 0.2060 & 0.2059 & 20.8\% & 97.1\% & 0.0028 & 0.0933 & 0.1769 & 0.1769 & 29.5\% & 99.3\% & 100.0\% \\
\cline{2-22}
 & \multirow{6}{*}{$n=5000$} & KM & 0.0099 & 0.0099 & 0.0038 & 0.0106 & 1.1\% & 100.0\% & 0.0950 & 0.0950 & 0.0100 & 0.0955 & 15.9\% & 100.0\% & 0.1338 & 0.1338 & 0.0129 & 0.1345 & 42.3\% & 100.0\% & 100.0\% \\
 &  & WCH & -0.0001 & 0.0038 & 0.0048 & 0.0048 & 0.4\% & 100.0\% & -0.0003 & 0.0144 & 0.0181 & 0.0180 & 2.4\% & 100.0\% & 0.0003 & 0.0182 & 0.0234 & 0.0234 & 5.7\% & 100.0\% & 100.0\% \\
 &  & WCM & 0.0000 & 0.0100 & 0.0125 & 0.0125 & 1.1\% & 100.0\% & -0.0010 & 0.0201 & 0.0254 & 0.0254 & 3.4\% & 100.0\% & -0.0005 & 0.0224 & 0.0295 & 0.0295 & 7.1\% & 100.0\% & 100.0\% \\
 &  & WFM & -0.0001 & 0.0039 & 0.0049 & 0.0049 & 0.4\% & 100.0\% & -0.0005 & 0.0176 & 0.0224 & 0.0224 & 2.9\% & 100.0\% & -0.0010 & 0.0312 & 0.0412 & 0.0412 & 9.9\% & 100.0\% & 100.0\% \\
 &  & WFH-2 & -0.0003 & 0.0049 & 0.0063 & 0.0063 & 0.5\% & 100.0\% & -0.0014 & 0.0209 & 0.0271 & 0.0272 & 3.5\% & 100.0\% & -0.0018 & 0.0266 & 0.0360 & 0.0361 & 8.4\% & 100.0\% & 100.0\% \\
 &  & WFM-2 & 0.0006 & 0.0486 & 0.0670 & 0.0670 & 5.2\% & 86.2\% & 0.0002 & 0.0448 & 0.0633 & 0.0633 & 7.5\% & 99.8\% & -0.0003 & 0.0361 & 0.0545 & 0.0545 & 11.4\% & 100.0\% & 100.0\% \\
\hline
\multirow{18}{*}{\shortstack{Moderate censoring\\(45\% events)}} & \multirow{6}{*}{$n=100$} & KM & 0.0043 & 0.0219 & 0.0267 & 0.0271 & 2.3\% & 100.0\% & 0.0598 & 0.0703 & 0.0603 & 0.0849 & 11.8\% & 100.0\% & 0.0883 & 0.0964 & 0.0731 & 0.1146 & 30.4\% & 100.0\% & 100.0\% \\
 &  & WCH & -0.0015 & 0.0234 & 0.0294 & 0.0295 & 2.5\% & 100.0\% & 0.0014 & 0.0577 & 0.0727 & 0.0727 & 9.6\% & 100.0\% & 0.0078 & 0.0631 & 0.0792 & 0.0795 & 19.9\% & 100.0\% & 100.0\% \\
 &  & WCM & -0.0015 & 0.0493 & 0.0618 & 0.0617 & 5.3\% & 86.4\% & 0.0018 & 0.0783 & 0.0985 & 0.0985 & 13.1\% & 100.0\% & 0.0057 & 0.0750 & 0.0997 & 0.0998 & 23.7\% & 99.9\% & 99.4\% \\
 &  & WFM & -0.0017 & 0.0236 & 0.0298 & 0.0298 & 2.5\% & 100.0\% & -0.0022 & 0.0696 & 0.0872 & 0.0872 & 11.6\% & 100.0\% & -0.0014 & 0.1001 & 0.1261 & 0.1260 & 31.6\% & 98.7\% & 100.0\% \\
 &  & WFH-2 & -0.0025 & 0.0252 & 0.0322 & 0.0323 & 2.7\% & 100.0\% & -0.0027 & 0.0686 & 0.0857 & 0.0857 & 11.5\% & 100.0\% & 0.0013 & 0.0764 & 0.0966 & 0.0965 & 24.1\% & 100.0\% & 100.0\% \\
 &  & WFM-2 & 0.0101 & 0.1309 & 0.1774 & 0.1776 & 14.0\% & 67.7\% & 0.0096 & 0.1227 & 0.1654 & 0.1656 & 20.5\% & 98.2\% & 0.0104 & 0.0985 & 0.1411 & 0.1414 & 31.1\% & 99.4\% & 100.0\% \\
\cline{2-22}
 & \multirow{6}{*}{$n=500$} & KM & 0.0053 & 0.0105 & 0.0118 & 0.0130 & 1.1\% & 100.0\% & 0.0599 & 0.0603 & 0.0270 & 0.0657 & 10.1\% & 100.0\% & 0.0835 & 0.0835 & 0.0323 & 0.0895 & 26.4\% & 100.0\% & 100.0\% \\
 &  & WCH & -0.0005 & 0.0105 & 0.0132 & 0.0132 & 1.1\% & 100.0\% & -0.0005 & 0.0260 & 0.0331 & 0.0331 & 4.3\% & 100.0\% & 0.0007 & 0.0282 & 0.0356 & 0.0356 & 8.9\% & 100.0\% & 100.0\% \\
 &  & WCM & 0.0007 & 0.0231 & 0.0290 & 0.0290 & 2.5\% & 98.8\% & -0.0013 & 0.0358 & 0.0447 & 0.0447 & 6.0\% & 100.0\% & -0.0028 & 0.0338 & 0.0428 & 0.0428 & 10.7\% & 100.0\% & 100.0\% \\
 &  & WFM & -0.0007 & 0.0106 & 0.0134 & 0.0134 & 1.1\% & 100.0\% & -0.0014 & 0.0310 & 0.0395 & 0.0395 & 5.2\% & 100.0\% & 0.0004 & 0.0451 & 0.0563 & 0.0563 & 14.2\% & 100.0\% & 100.0\% \\
 &  & WFH-2 & -0.0012 & 0.0111 & 0.0139 & 0.0140 & 1.2\% & 100.0\% & -0.0042 & 0.0311 & 0.0389 & 0.0391 & 5.2\% & 100.0\% & -0.0035 & 0.0334 & 0.0423 & 0.0424 & 10.5\% & 100.0\% & 100.0\% \\
 &  & WFM-2 & -0.0051 & 0.0592 & 0.0739 & 0.0740 & 6.3\% & 83.6\% & -0.0058 & 0.0547 & 0.0683 & 0.0685 & 9.1\% & 100.0\% & -0.0040 & 0.0422 & 0.0542 & 0.0543 & 13.3\% & 100.0\% & 100.0\% \\
\cline{2-22}
 & \multirow{6}{*}{$n=5000$} & KM & 0.0059 & 0.0061 & 0.0037 & 0.0069 & 0.7\% & 100.0\% & 0.0609 & 0.0609 & 0.0086 & 0.0615 & 10.2\% & 100.0\% & 0.0858 & 0.0858 & 0.0102 & 0.0865 & 27.1\% & 100.0\% & 100.0\% \\
 &  & WCH & 0.0001 & 0.0033 & 0.0041 & 0.0041 & 0.3\% & 100.0\% & 0.0005 & 0.0084 & 0.0107 & 0.0107 & 1.4\% & 100.0\% & 0.0003 & 0.0094 & 0.0118 & 0.0118 & 3.0\% & 100.0\% & 100.0\% \\
 &  & WCM & -0.0002 & 0.0070 & 0.0088 & 0.0088 & 0.8\% & 100.0\% & 0.0005 & 0.0110 & 0.0138 & 0.0138 & 1.8\% & 100.0\% & 0.0002 & 0.0107 & 0.0134 & 0.0134 & 3.4\% & 100.0\% & 100.0\% \\
 &  & WFM & 0.0001 & 0.0033 & 0.0041 & 0.0041 & 0.4\% & 100.0\% & 0.0005 & 0.0102 & 0.0126 & 0.0126 & 1.7\% & 100.0\% & 0.0002 & 0.0147 & 0.0182 & 0.0182 & 4.6\% & 100.0\% & 100.0\% \\
 &  & WFH-2 & 0.0001 & 0.0034 & 0.0043 & 0.0043 & 0.4\% & 100.0\% & 0.0006 & 0.0100 & 0.0126 & 0.0126 & 1.7\% & 100.0\% & 0.0003 & 0.0116 & 0.0144 & 0.0144 & 3.7\% & 100.0\% & 100.0\% \\
 &  & WFM-2 & 0.0005 & 0.0189 & 0.0236 & 0.0236 & 2.0\% & 99.6\% & 0.0009 & 0.0175 & 0.0223 & 0.0223 & 2.9\% & 100.0\% & 0.0006 & 0.0142 & 0.0181 & 0.0181 & 4.5\% & 100.0\% & 100.0\% \\
\hline
\multirow{18}{*}{\shortstack{Lower censoring\\(60\% events)}} & \multirow{6}{*}{$n=100$} & KM & 0.0023 & 0.0202 & 0.0258 & 0.0259 & 2.2\% & 100.0\% & 0.0408 & 0.0574 & 0.0574 & 0.0704 & 9.6\% & 100.0\% & 0.0558 & 0.0683 & 0.0617 & 0.0832 & 21.6\% & 100.0\% & 100.0\% \\
 &  & WCH & -0.0015 & 0.0213 & 0.0274 & 0.0275 & 2.3\% & 100.0\% & 0.0022 & 0.0507 & 0.0635 & 0.0635 & 8.5\% & 100.0\% & 0.0019 & 0.0493 & 0.0621 & 0.0621 & 15.6\% & 100.0\% & 100.0\% \\
 &  & WCM & -0.0020 & 0.0383 & 0.0480 & 0.0480 & 4.1\% & 92.5\% & 0.0009 & 0.0615 & 0.0765 & 0.0764 & 10.3\% & 100.0\% & -0.0008 & 0.0530 & 0.0669 & 0.0668 & 16.8\% & 100.0\% & 100.0\% \\
 &  & WFM & -0.0015 & 0.0215 & 0.0278 & 0.0278 & 2.3\% & 100.0\% & 0.0016 & 0.0566 & 0.0707 & 0.0707 & 9.5\% & 100.0\% & 0.0010 & 0.0698 & 0.0882 & 0.0881 & 22.1\% & 100.0\% & 100.0\% \\
 &  & WFH-2 & -0.0020 & 0.0217 & 0.0282 & 0.0283 & 2.3\% & 100.0\% & -0.0011 & 0.0547 & 0.0685 & 0.0685 & 9.1\% & 100.0\% & -0.0023 & 0.0545 & 0.0687 & 0.0687 & 17.2\% & 100.0\% & 100.0\% \\
 &  & WFM-2 & -0.0047 & 0.0775 & 0.0988 & 0.0989 & 8.3\% & 77.3\% & -0.0016 & 0.0778 & 0.0992 & 0.0992 & 13.0\% & 99.8\% & -0.0022 & 0.0625 & 0.0798 & 0.0798 & 19.7\% & 100.0\% & 100.0\% \\
\cline{2-22}
 & \multirow{6}{*}{$n=500$} & KM & 0.0038 & 0.0094 & 0.0111 & 0.0117 & 1.0\% & 100.0\% & 0.0392 & 0.0405 & 0.0254 & 0.0467 & 6.8\% & 100.0\% & 0.0541 & 0.0547 & 0.0260 & 0.0600 & 17.3\% & 100.0\% & 100.0\% \\
 &  & WCH & 0.0002 & 0.0095 & 0.0118 & 0.0118 & 1.0\% & 100.0\% & 0.0001 & 0.0221 & 0.0277 & 0.0277 & 3.7\% & 100.0\% & -0.0005 & 0.0209 & 0.0267 & 0.0267 & 6.6\% & 100.0\% & 100.0\% \\
 &  & WCM & -0.0001 & 0.0178 & 0.0224 & 0.0224 & 1.9\% & 99.9\% & -0.0016 & 0.0284 & 0.0350 & 0.0351 & 4.7\% & 100.0\% & -0.0017 & 0.0242 & 0.0308 & 0.0308 & 7.6\% & 100.0\% & 100.0\% \\
 &  & WFM & 0.0002 & 0.0095 & 0.0118 & 0.0118 & 1.0\% & 100.0\% & 0.0005 & 0.0244 & 0.0305 & 0.0305 & 4.1\% & 100.0\% & 0.0004 & 0.0301 & 0.0380 & 0.0379 & 9.5\% & 100.0\% & 100.0\% \\
 &  & WFH-2 & -0.0000 & 0.0097 & 0.0121 & 0.0121 & 1.0\% & 100.0\% & -0.0007 & 0.0240 & 0.0298 & 0.0298 & 4.0\% & 100.0\% & -0.0014 & 0.0237 & 0.0299 & 0.0300 & 7.5\% & 100.0\% & 100.0\% \\
 &  & WFM-2 & -0.0021 & 0.0364 & 0.0456 & 0.0456 & 3.9\% & 93.0\% & -0.0017 & 0.0361 & 0.0446 & 0.0447 & 6.0\% & 100.0\% & -0.0019 & 0.0285 & 0.0357 & 0.0357 & 9.0\% & 100.0\% & 100.0\% \\
\cline{2-22}
 & \multirow{6}{*}{$n=5000$} & KM & 0.0036 & 0.0042 & 0.0036 & 0.0051 & 0.5\% & 100.0\% & 0.0389 & 0.0389 & 0.0078 & 0.0397 & 6.5\% & 100.0\% & 0.0549 & 0.0549 & 0.0085 & 0.0556 & 17.3\% & 100.0\% & 100.0\% \\
 &  & WCH & -0.0000 & 0.0030 & 0.0039 & 0.0039 & 0.3\% & 100.0\% & -0.0003 & 0.0067 & 0.0085 & 0.0085 & 1.1\% & 100.0\% & -0.0004 & 0.0068 & 0.0086 & 0.0086 & 2.1\% & 100.0\% & 100.0\% \\
 &  & WCM & 0.0001 & 0.0058 & 0.0071 & 0.0071 & 0.6\% & 100.0\% & -0.0001 & 0.0085 & 0.0107 & 0.0107 & 1.4\% & 100.0\% & -0.0002 & 0.0076 & 0.0096 & 0.0096 & 2.4\% & 100.0\% & 100.0\% \\
 &  & WFM & -0.0000 & 0.0031 & 0.0039 & 0.0039 & 0.3\% & 100.0\% & -0.0004 & 0.0077 & 0.0096 & 0.0096 & 1.3\% & 100.0\% & -0.0007 & 0.0096 & 0.0120 & 0.0120 & 3.0\% & 100.0\% & 100.0\% \\
 &  & WFH-2 & 0.0000 & 0.0031 & 0.0040 & 0.0040 & 0.3\% & 100.0\% & -0.0001 & 0.0074 & 0.0094 & 0.0094 & 1.2\% & 100.0\% & -0.0001 & 0.0081 & 0.0100 & 0.0100 & 2.5\% & 100.0\% & 100.0\% \\
 &  & WFM-2 & 0.0008 & 0.0113 & 0.0140 & 0.0140 & 1.2\% & 100.0\% & 0.0005 & 0.0111 & 0.0139 & 0.0139 & 1.9\% & 100.0\% & 0.0002 & 0.0094 & 0.0118 & 0.0118 & 3.0\% & 100.0\% & 100.0\% \\
\hline
\bottomrule
\end{tabular}
\end{adjustbox}
\end{table}
\endgroup

\clearpage

\begingroup
\setlength{\intextsep}{0pt}
\begin{table}[H]
\centering
\caption{\sh{Simulation results for Setting II at $n=100,500,5000$. MARE is mean absolute relative error; range is the percentage in $[0,1]$; Mono. assesses the three evaluation times.}}\label{tab:weibull-admin}
\fontsize{8}{8.75}\selectfont
\setlength{\tabcolsep}{3.2pt}
\renewcommand{\arraystretch}{0.94}
\begin{adjustbox}{max width=\linewidth}
\begin{tabular}{c|c|c|cccccc|cccccc|cccccc|c}
\toprule
\multirow{2}{*}{Scheme} & \multirow{2}{*}{$n$} & \multirow{2}{*}{Method} & \multicolumn{6}{c|}{$t=1,\ S(t)=0.9351$} & \multicolumn{6}{c|}{$t=3,\ S(t)=0.5985$} & \multicolumn{6}{c|}{$t=5,\ S(t)=0.3166$} & \multirow{2}{*}{Mono.} \\
\cline{4-21}
 & & & bias & MAE & sd & rmse & MARE & range & bias & MAE & sd & rmse & MARE & range & bias & MAE & sd & rmse & MARE & range & \\
\hline
\multirow{18}{*}{\shortstack{Higher censoring\\(23\% events)}} & \multirow{6}{*}{$n=100$} & KM & 0.0110 & 0.0233 & 0.0262 & 0.0284 & 2.5\% & 100.0\% & 0.0951 & 0.1013 & 0.0709 & 0.1186 & 16.9\% & 100.0\% & 0.1342 & 0.1404 & 0.0942 & 0.1640 & 44.3\% & 100.0\% & 100.0\% \\
 &  & WCH & 0.0017 & 0.0262 & 0.0329 & 0.0329 & 2.8\% & 100.0\% & 0.0064 & 0.0863 & 0.1084 & 0.1086 & 14.4\% & 100.0\% & 0.0194 & 0.0980 & 0.1195 & 0.1210 & 31.0\% & 100.0\% & 100.0\% \\
 &  & WCM & 0.0038 & 0.0755 & 0.0942 & 0.0942 & 8.1\% & 73.0\% & -0.0018 & 0.1357 & 0.1891 & 0.1890 & 22.7\% & 97.1\% & -0.0022 & 0.1158 & 0.1791 & 0.1790 & 36.6\% & 99.5\% & 90.4\% \\
 &  & WFM & 0.0016 & 0.0267 & 0.0334 & 0.0334 & 2.9\% & 100.0\% & -0.0032 & 0.1116 & 0.1588 & 0.1587 & 18.6\% & 99.1\% & -0.0013 & 0.1835 & 0.2685 & 0.2683 & 58.0\% & 92.3\% & 100.0\% \\
 &  & WFH-2 & -0.0242 & 0.0427 & 0.0547 & 0.0598 & 4.6\% & 100.0\% & -0.1420 & 0.1789 & 0.1605 & 0.2142 & 29.9\% & 100.0\% & -0.2024 & 0.2170 & 0.1146 & 0.2326 & 68.6\% & 100.0\% & 100.0\% \\
 &  & WFM-2 & -0.2202 & 0.3069 & 0.2934 & 0.3668 & 32.8\% & 88.9\% & -0.2250 & 0.2952 & 0.2617 & 0.3451 & 49.3\% & 97.3\% & -0.2231 & 0.2421 & 0.1286 & 0.2574 & 76.5\% & 99.4\% & 100.0\% \\
\cline{2-22}
 & \multirow{6}{*}{$n=500$} & KM & 0.0100 & 0.0128 & 0.0118 & 0.0155 & 1.4\% & 100.0\% & 0.0952 & 0.0952 & 0.0326 & 0.1007 & 15.9\% & 100.0\% & 0.1358 & 0.1358 & 0.0420 & 0.1421 & 42.9\% & 100.0\% & 100.0\% \\
 &  & WCH & 0.0002 & 0.0118 & 0.0149 & 0.0149 & 1.3\% & 100.0\% & 0.0022 & 0.0434 & 0.0553 & 0.0553 & 7.3\% & 100.0\% & 0.0105 & 0.0530 & 0.0667 & 0.0675 & 16.8\% & 100.0\% & 100.0\% \\
 &  & WCM & -0.0034 & 0.0319 & 0.0397 & 0.0399 & 3.4\% & 95.5\% & -0.0045 & 0.0654 & 0.0844 & 0.0845 & 10.9\% & 99.9\% & 0.0014 & 0.0631 & 0.1003 & 0.1002 & 19.9\% & 99.7\% & 98.4\% \\
 &  & WFM & 0.0003 & 0.0120 & 0.0151 & 0.0151 & 1.3\% & 100.0\% & 0.0013 & 0.0531 & 0.0703 & 0.0703 & 8.9\% & 100.0\% & 0.0047 & 0.0964 & 0.1397 & 0.1397 & 30.4\% & 97.8\% & 100.0\% \\
 &  & WFH-2 & -0.0216 & 0.0253 & 0.0240 & 0.0323 & 2.7\% & 100.0\% & -0.1279 & 0.1363 & 0.0855 & 0.1538 & 22.8\% & 100.0\% & -0.2019 & 0.2054 & 0.0670 & 0.2127 & 64.9\% & 100.0\% & 100.0\% \\
 &  & WFM-2 & -0.2304 & 0.2545 & 0.1600 & 0.2804 & 27.2\% & 95.7\% & -0.2293 & 0.2515 & 0.1455 & 0.2716 & 42.0\% & 99.1\% & -0.2259 & 0.2329 & 0.0827 & 0.2406 & 73.6\% & 99.8\% & 100.0\% \\
\cline{2-22}
 & \multirow{6}{*}{$n=5000$} & KM & 0.0097 & 0.0097 & 0.0039 & 0.0105 & 1.0\% & 100.0\% & 0.0953 & 0.0953 & 0.0100 & 0.0959 & 15.9\% & 100.0\% & 0.1346 & 0.1346 & 0.0134 & 0.1353 & 42.5\% & 100.0\% & 100.0\% \\
 &  & WCH & -0.0004 & 0.0039 & 0.0048 & 0.0048 & 0.4\% & 100.0\% & -0.0005 & 0.0142 & 0.0176 & 0.0176 & 2.4\% & 100.0\% & 0.0023 & 0.0195 & 0.0244 & 0.0245 & 6.2\% & 100.0\% & 100.0\% \\
 &  & WCM & 0.0001 & 0.0104 & 0.0130 & 0.0130 & 1.1\% & 100.0\% & -0.0012 & 0.0209 & 0.0260 & 0.0261 & 3.5\% & 100.0\% & -0.0013 & 0.0215 & 0.0285 & 0.0285 & 6.8\% & 100.0\% & 100.0\% \\
 &  & WFM & -0.0004 & 0.0039 & 0.0049 & 0.0049 & 0.4\% & 100.0\% & -0.0008 & 0.0179 & 0.0222 & 0.0222 & 3.0\% & 100.0\% & 0.0020 & 0.0326 & 0.0410 & 0.0410 & 10.3\% & 100.0\% & 100.0\% \\
 &  & WFH-2 & -0.0198 & 0.0198 & 0.0078 & 0.0213 & 2.1\% & 100.0\% & -0.1199 & 0.1199 & 0.0302 & 0.1236 & 20.0\% & 100.0\% & -0.1984 & 0.1984 & 0.0267 & 0.2002 & 62.7\% & 100.0\% & 100.0\% \\
 &  & WFM-2 & -0.2272 & 0.2273 & 0.0469 & 0.2320 & 24.3\% & 100.0\% & -0.2276 & 0.2276 & 0.0413 & 0.2314 & 38.0\% & 100.0\% & -0.2248 & 0.2248 & 0.0240 & 0.2261 & 71.0\% & 100.0\% & 100.0\% \\
\hline
\multirow{18}{*}{\shortstack{Moderate censoring\\(37\% events)}} & \multirow{6}{*}{$n=100$} & KM & 0.0065 & 0.0208 & 0.0250 & 0.0258 & 2.2\% & 100.0\% & 0.0574 & 0.0706 & 0.0636 & 0.0857 & 11.8\% & 100.0\% & 0.0803 & 0.0898 & 0.0723 & 0.1080 & 28.4\% & 100.0\% & 100.0\% \\
 &  & WCH & 0.0009 & 0.0220 & 0.0278 & 0.0278 & 2.4\% & 100.0\% & -0.0016 & 0.0609 & 0.0753 & 0.0752 & 10.2\% & 100.0\% & 0.0033 & 0.0622 & 0.0785 & 0.0785 & 19.6\% & 100.0\% & 100.0\% \\
 &  & WCM & 0.0002 & 0.0474 & 0.0591 & 0.0591 & 5.1\% & 87.6\% & -0.0054 & 0.0803 & 0.0992 & 0.0993 & 13.4\% & 100.0\% & -0.0061 & 0.0725 & 0.0929 & 0.0931 & 22.9\% & 100.0\% & 99.4\% \\
 &  & WFM & 0.0008 & 0.0226 & 0.0283 & 0.0283 & 2.4\% & 100.0\% & -0.0035 & 0.0713 & 0.0892 & 0.0892 & 11.9\% & 100.0\% & 0.0002 & 0.0983 & 0.1239 & 0.1238 & 31.1\% & 99.2\% & 100.0\% \\
 &  & WFH-2 & -0.0196 & 0.0331 & 0.0385 & 0.0432 & 3.5\% & 100.0\% & -0.1291 & 0.1406 & 0.1031 & 0.1651 & 23.5\% & 100.0\% & -0.2050 & 0.2068 & 0.0740 & 0.2179 & 65.3\% & 100.0\% & 100.0\% \\
 &  & WFM-2 & -0.2276 & 0.2361 & 0.1354 & 0.2648 & 25.2\% & 97.5\% & -0.2319 & 0.2386 & 0.1198 & 0.2610 & 39.9\% & 100.0\% & -0.2283 & 0.2301 & 0.0669 & 0.2379 & 72.7\% & 100.0\% & 100.0\% \\
\cline{2-22}
 & \multirow{6}{*}{$n=500$} & KM & 0.0056 & 0.0104 & 0.0117 & 0.0129 & 1.1\% & 100.0\% & 0.0601 & 0.0604 & 0.0270 & 0.0659 & 10.1\% & 100.0\% & 0.0857 & 0.0857 & 0.0318 & 0.0914 & 27.1\% & 100.0\% & 100.0\% \\
 &  & WCH & -0.0002 & 0.0103 & 0.0130 & 0.0130 & 1.1\% & 100.0\% & -0.0006 & 0.0264 & 0.0328 & 0.0327 & 4.4\% & 100.0\% & 0.0010 & 0.0287 & 0.0360 & 0.0360 & 9.1\% & 100.0\% & 100.0\% \\
 &  & WCM & 0.0007 & 0.0223 & 0.0281 & 0.0281 & 2.4\% & 98.4\% & 0.0004 & 0.0367 & 0.0457 & 0.0456 & 6.1\% & 100.0\% & 0.0013 & 0.0326 & 0.0410 & 0.0410 & 10.3\% & 100.0\% & 100.0\% \\
 &  & WFM & -0.0002 & 0.0104 & 0.0131 & 0.0131 & 1.1\% & 100.0\% & -0.0016 & 0.0311 & 0.0388 & 0.0388 & 5.2\% & 100.0\% & -0.0019 & 0.0461 & 0.0577 & 0.0577 & 14.5\% & 100.0\% & 100.0\% \\
 &  & WFH-2 & -0.0193 & 0.0214 & 0.0180 & 0.0264 & 2.3\% & 100.0\% & -0.1180 & 0.1180 & 0.0457 & 0.1266 & 19.7\% & 100.0\% & -0.1981 & 0.1981 & 0.0347 & 0.2011 & 62.6\% & 100.0\% & 100.0\% \\
 &  & WFM-2 & -0.2223 & 0.2224 & 0.0650 & 0.2316 & 23.8\% & 100.0\% & -0.2237 & 0.2237 & 0.0555 & 0.2305 & 37.4\% & 100.0\% & -0.2240 & 0.2240 & 0.0302 & 0.2260 & 70.7\% & 100.0\% & 100.0\% \\
\cline{2-22}
 & \multirow{6}{*}{$n=5000$} & KM & 0.0058 & 0.0060 & 0.0037 & 0.0069 & 0.6\% & 100.0\% & 0.0607 & 0.0607 & 0.0085 & 0.0613 & 10.1\% & 100.0\% & 0.0857 & 0.0857 & 0.0101 & 0.0863 & 27.1\% & 100.0\% & 100.0\% \\
 &  & WCH & -0.0000 & 0.0033 & 0.0041 & 0.0041 & 0.3\% & 100.0\% & -0.0002 & 0.0081 & 0.0103 & 0.0103 & 1.3\% & 100.0\% & 0.0006 & 0.0093 & 0.0119 & 0.0119 & 2.9\% & 100.0\% & 100.0\% \\
 &  & WCM & -0.0003 & 0.0069 & 0.0086 & 0.0086 & 0.7\% & 100.0\% & -0.0002 & 0.0111 & 0.0141 & 0.0141 & 1.9\% & 100.0\% & 0.0002 & 0.0101 & 0.0127 & 0.0127 & 3.2\% & 100.0\% & 100.0\% \\
 &  & WFM & -0.0000 & 0.0033 & 0.0041 & 0.0041 & 0.4\% & 100.0\% & -0.0002 & 0.0095 & 0.0121 & 0.0121 & 1.6\% & 100.0\% & 0.0005 & 0.0149 & 0.0189 & 0.0189 & 4.7\% & 100.0\% & 100.0\% \\
 &  & WFH-2 & -0.0189 & 0.0189 & 0.0055 & 0.0197 & 2.0\% & 100.0\% & -0.1172 & 0.1172 & 0.0142 & 0.1181 & 19.6\% & 100.0\% & -0.1982 & 0.1982 & 0.0110 & 0.1985 & 62.6\% & 100.0\% & 100.0\% \\
 &  & WFM-2 & -0.2254 & 0.2254 & 0.0200 & 0.2262 & 24.1\% & 100.0\% & -0.2256 & 0.2256 & 0.0170 & 0.2262 & 37.7\% & 100.0\% & -0.2248 & 0.2248 & 0.0090 & 0.2250 & 71.0\% & 100.0\% & 100.0\% \\
\hline
\multirow{18}{*}{\shortstack{Lower censoring\\(48\% events)}} & \multirow{6}{*}{$n=100$} & KM & 0.0032 & 0.0211 & 0.0259 & 0.0261 & 2.3\% & 100.0\% & 0.0384 & 0.0554 & 0.0566 & 0.0684 & 9.3\% & 100.0\% & 0.0552 & 0.0684 & 0.0630 & 0.0838 & 21.6\% & 100.0\% & 100.0\% \\
 &  & WCH & -0.0006 & 0.0221 & 0.0278 & 0.0278 & 2.4\% & 100.0\% & 0.0002 & 0.0498 & 0.0619 & 0.0619 & 8.3\% & 100.0\% & 0.0002 & 0.0516 & 0.0644 & 0.0644 & 16.3\% & 100.0\% & 100.0\% \\
 &  & WCM & -0.0023 & 0.0407 & 0.0510 & 0.0510 & 4.3\% & 90.1\% & 0.0021 & 0.0631 & 0.0774 & 0.0774 & 10.5\% & 100.0\% & -0.0002 & 0.0561 & 0.0695 & 0.0695 & 17.7\% & 100.0\% & 100.0\% \\
 &  & WFM & -0.0005 & 0.0222 & 0.0277 & 0.0277 & 2.4\% & 100.0\% & -0.0012 & 0.0555 & 0.0695 & 0.0694 & 9.3\% & 100.0\% & -0.0036 & 0.0727 & 0.0913 & 0.0914 & 23.0\% & 99.9\% & 100.0\% \\
 &  & WFH-2 & -0.0199 & 0.0323 & 0.0370 & 0.0420 & 3.5\% & 100.0\% & -0.1163 & 0.1198 & 0.0770 & 0.1394 & 20.0\% & 100.0\% & -0.1972 & 0.1972 & 0.0536 & 0.2044 & 62.3\% & 100.0\% & 100.0\% \\
 &  & WFM-2 & -0.2201 & 0.2211 & 0.0969 & 0.2404 & 23.6\% & 99.6\% & -0.2207 & 0.2213 & 0.0830 & 0.2358 & 37.0\% & 100.0\% & -0.2231 & 0.2231 & 0.0438 & 0.2274 & 70.5\% & 100.0\% & 100.0\% \\
\cline{2-22}
 & \multirow{6}{*}{$n=500$} & KM & 0.0037 & 0.0094 & 0.0111 & 0.0117 & 1.0\% & 100.0\% & 0.0391 & 0.0404 & 0.0250 & 0.0464 & 6.8\% & 100.0\% & 0.0554 & 0.0557 & 0.0274 & 0.0618 & 17.6\% & 100.0\% & 100.0\% \\
 &  & WCH & 0.0001 & 0.0093 & 0.0118 & 0.0118 & 1.0\% & 100.0\% & 0.0003 & 0.0220 & 0.0273 & 0.0273 & 3.7\% & 100.0\% & 0.0009 & 0.0221 & 0.0277 & 0.0277 & 7.0\% & 100.0\% & 100.0\% \\
 &  & WCM & -0.0005 & 0.0174 & 0.0215 & 0.0215 & 1.9\% & 99.8\% & -0.0011 & 0.0266 & 0.0331 & 0.0331 & 4.4\% & 100.0\% & -0.0005 & 0.0244 & 0.0305 & 0.0305 & 7.7\% & 100.0\% & 100.0\% \\
 &  & WFM & 0.0002 & 0.0094 & 0.0119 & 0.0119 & 1.0\% & 100.0\% & 0.0006 & 0.0248 & 0.0312 & 0.0312 & 4.1\% & 100.0\% & 0.0017 & 0.0316 & 0.0399 & 0.0399 & 10.0\% & 100.0\% & 100.0\% \\
 &  & WFH-2 & -0.0189 & 0.0204 & 0.0157 & 0.0246 & 2.2\% & 100.0\% & -0.1168 & 0.1168 & 0.0353 & 0.1221 & 19.5\% & 100.0\% & -0.1975 & 0.1975 & 0.0250 & 0.1991 & 62.4\% & 100.0\% & 100.0\% \\
 &  & WFM-2 & -0.2258 & 0.2258 & 0.0427 & 0.2298 & 24.1\% & 100.0\% & -0.2254 & 0.2254 & 0.0372 & 0.2284 & 37.7\% & 100.0\% & -0.2243 & 0.2243 & 0.0206 & 0.2252 & 70.8\% & 100.0\% & 100.0\% \\
\cline{2-22}
 & \multirow{6}{*}{$n=5000$} & KM & 0.0034 & 0.0042 & 0.0037 & 0.0051 & 0.5\% & 100.0\% & 0.0390 & 0.0390 & 0.0078 & 0.0398 & 6.5\% & 100.0\% & 0.0555 & 0.0555 & 0.0086 & 0.0562 & 17.5\% & 100.0\% & 100.0\% \\
 &  & WCH & -0.0002 & 0.0032 & 0.0040 & 0.0040 & 0.3\% & 100.0\% & -0.0003 & 0.0067 & 0.0084 & 0.0084 & 1.1\% & 100.0\% & 0.0001 & 0.0069 & 0.0086 & 0.0086 & 2.2\% & 100.0\% & 100.0\% \\
 &  & WCM & -0.0005 & 0.0056 & 0.0070 & 0.0070 & 0.6\% & 100.0\% & -0.0007 & 0.0085 & 0.0107 & 0.0108 & 1.4\% & 100.0\% & -0.0002 & 0.0079 & 0.0098 & 0.0098 & 2.5\% & 100.0\% & 100.0\% \\
 &  & WFM & -0.0002 & 0.0032 & 0.0040 & 0.0040 & 0.3\% & 100.0\% & -0.0002 & 0.0074 & 0.0094 & 0.0094 & 1.2\% & 100.0\% & 0.0004 & 0.0096 & 0.0122 & 0.0122 & 3.0\% & 100.0\% & 100.0\% \\
 &  & WFH-2 & -0.0191 & 0.0191 & 0.0052 & 0.0198 & 2.0\% & 100.0\% & -0.1168 & 0.1168 & 0.0107 & 0.1173 & 19.5\% & 100.0\% & -0.1979 & 0.1979 & 0.0075 & 0.1980 & 62.5\% & 100.0\% & 100.0\% \\
 &  & WFM-2 & -0.2252 & 0.2252 & 0.0133 & 0.2255 & 24.1\% & 100.0\% & -0.2251 & 0.2251 & 0.0115 & 0.2254 & 37.6\% & 100.0\% & -0.2246 & 0.2246 & 0.0062 & 0.2247 & 70.9\% & 100.0\% & 100.0\% \\
\hline
\bottomrule
\end{tabular}
\end{adjustbox}
\end{table}
\endgroup

\end{landscape}
\restoregeometry
\section{Application}

\sh{To illustrate the practical implications of IPCW estimator structure, we analyzed the publicly available Women's Interagency HIV Study (WIHS) data distributed with the \texttt{randomForestSRC} package \citep{bacon2005wihs,ishwaran2025randomforestsrc}. The data include 1,164 women who were alive, infected with HIV, and free of clinical AIDS in December 1995. Participants were followed until HAART initiation, AIDS diagnosis or death, or administrative censoring in September 2006. We treated AIDS diagnosis or death before HAART initiation as the event and treated HAART initiation and other loss to follow-up as censoring. There were 359 observed AIDS/death events, 679 HAART initiations, and 126 administratively or otherwise censored observations}.

\sh{The resulting curve can be interpreted as AIDS/death-free survival, assuming that censoring is conditionally independent of the latent event time given the baseline covariates. We estimated $G(t\mid L)$ using a Weibull regression model containing age, history of injection drug use, race, and CD4 nadir prior to baseline. All IPCW estimators used this same fitted censoring distribution and differed only in how the weights entered the survival estimator. }

\begin{figure}[htbp]
    \centering
    \includegraphics[width=0.97\textwidth]{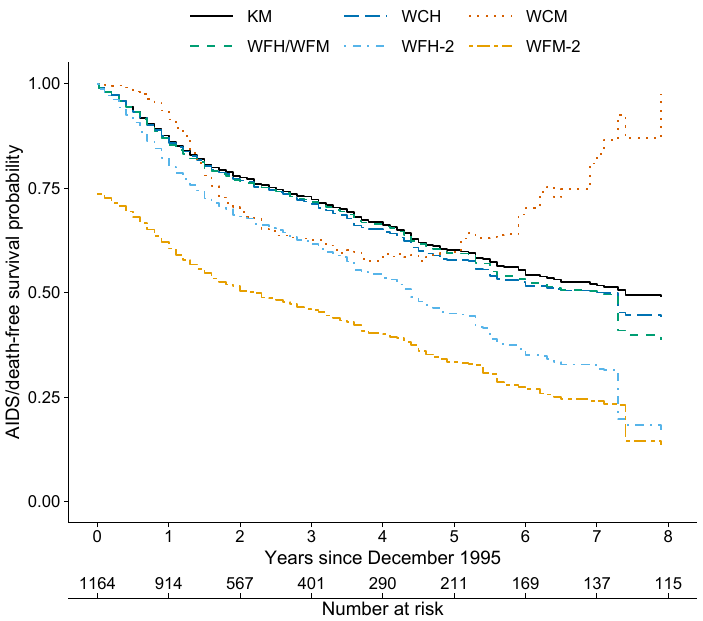}
    \caption{\sh{Estimated AIDS/death-free survival curves in the WIHS data, with observed numbers at risk shared by all methods. WFM-2 starts at $\bar w=0.736$.}}
    \label{fig:wihs}
\end{figure}

\sh{Figure~\ref{fig:wihs} displays the estimated curves through eight years. KM, WCH, and WFH/WFM were similar early in follow-up but separated as the risk set contracted and censoring accumulated. The lower WCH estimate relative to KM is consistent with a potential informative-censoring mechanism in which individuals at greater risk of AIDS or death are more likely to be censored, for example through earlier HAART initiation. The other three constructions diverged more substantially during later follow-up. WCM became strongly nonmonotone and increased to 0.975 by 7.9 years, whereas WFH-2 and WFM-2 produced substantially lower estimates, consistent with their additional sensitivity to bounded follow-up support. In contrast, WCH remained monotone and bounded while using time-specific risk-set normalization. These results parallel the simulation findings: discrepancies among the constructions become most visible after censoring has accumulated and the observed risk set has become small. They also demonstrate why using a common fitted censoring model does not make the resulting IPCW survival estimators interchangeable.}

\begin{figure}[htbp]
\centering
\includegraphics[width=0.97\textwidth]{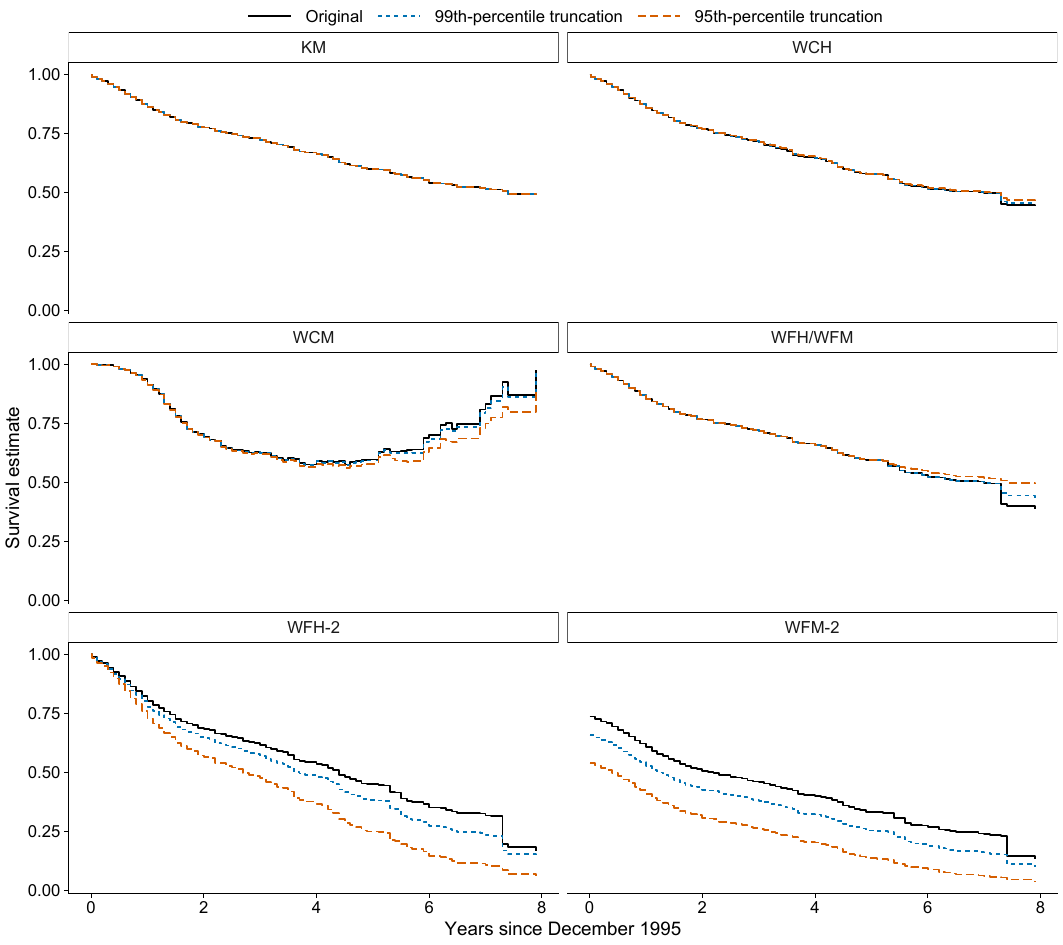}
\caption{\sh{WIHS sensitivity analysis comparing the primary estimates with estimates using weights truncated at the 99th and 95th percentiles.}}
\label{fig:wihs-sensitivity}
\end{figure}

\sh{To assess sensitivity to large weights, we compared the primary estimates with estimates obtained by truncating weights at the empirical 99th and 95th percentiles, replacing weights above each threshold with the threshold value. For WCH and WCM, thresholds were calculated within the risk set at each event or evaluation time, with the same truncated weights used in the numerator and denominator of WCH. For fixed-weight estimators, one threshold was calculated from the positive observed-failure weights across the dataset and applied before constructing each estimator. Figure~\ref{fig:wihs-sensitivity} shows the resulting curves. Estimators using time-changing weights were generally less sensitive to weight truncation than those using fixed weights, and the main differences among estimator constructions persisted.}

\FloatBarrier

\section{Discussion}

In this work, we examined a class of inverse probability of censoring weighting (IPCW) estimators for the survival function and demonstrated that how inverse probability weights are incorporated into survival estimation procedures can materially affect finite-sample behavior, even under a common censoring model. While IPCW is often described as a unified framework, our results show that different constructions, particularly margin-based versus hazard-based formulations and time-fixed versus time-varying weighting schemes, can lead to very different finite-sample behavior, even when the censoring model is correctly specified.
Across the simulation settings, the weight-changing hazard (WCH) estimator had the most favorable bias-variance tradeoff at later evaluation times and always produced a valid survival curve; in the application, it remained monotone and bounded where the alternative constructions did not.

The favorable performance of the WCH estimator {appears closely related to the combination of its time-varying weighting scheme and structural properties}. By embedding time-varying inverse censoring weights within a product-limit framework, WCH preserves the defining characteristics of survival functions, including monotonicity and boundedness, while leveraging risk-set-based normalization at each time point. This normalization plays a critical role in stabilizing estimation, particularly in settings with heavy censoring or limited numbers of observed events, where effective sample sizes shrink over time.
{The limitations of the alternative estimators arise because inverse weighting alone does not ensure either stability or structural validity. Estimators without dynamic risk-set normalization can be highly sensitive to a small number of large weighted contributions, while pointwise constructions need not preserve monotonicity or boundedness across time. A product-limit form alone is also insufficient when its weighted hazard increments are not properly normalized. In addition, estimators that rely exclusively on weighted observed failures may require event-time support beyond the evaluation time and can therefore target support-restricted quantities under administrative censoring. These findings underscore that reliable IPCW survival estimation depends jointly on the weighting scheme, normalization, and estimator construction.}

The differences among IPCW estimators are most pronounced in scenarios that are commonly encountered in biomedical studies, including settings with heavy or covariate-dependent censoring, \sh{administrative censoring}, limited sample size, small event counts, and inference focused on later follow-up times.
In these cases, instability in inverse weights and contraction of the risk set can amplify discrepancies between estimators, leading not only to increased variance but also to violations of fundamental properties of survival functions and identification of unintended target quantities.
As demonstrated in both our simulations and application, these effects can translate into materially different survival estimates, with potential implications for clinical interpretation and decision-making.

This work also has several limitations. First, we focused on settings in which the censoring distribution is correctly specified, allowing us to isolate differences attributable to estimator structure. \sh{The simulations additionally used the true censoring probabilities and therefore did not incorporate the uncertainty associated with estimating the censoring distribution.} In practice, misspecification of the censoring model may further impact estimator performance and interact with the structural properties studied here. Second, our analysis was restricted to right-censored data with baseline covariates, and extensions to settings with time-dependent covariates or more complex censoring mechanisms warrant further investigation. \sh{This limitation is particularly relevant to the WIHS application, in which HAART initiation was treated as censoring but may depend on post-baseline clinical information. The application should therefore be viewed mainly as an illustration of differences among estimator constructions.} Finally, while our simulations were designed to capture a range of realistic scenarios, additional studies may be needed to assess performance under alternative data-generating mechanisms. \sh{The present analysis also did not provide uncertainty quantification for the estimated survival curves.} Several directions for future work follow naturally from our findings. For example, extending the hazard-based IPCW framework to incorporate flexible, data-adaptive estimation of the censoring distribution may further improve robustness in high-dimensional settings. More broadly, the distinction between estimator structure and weight estimation highlighted in this work suggests that similar considerations may arise in other areas of causal and survival analysis, where inverse weighting plays a central role.

\section*{Statements and Declarations}
\subsection*{Author contributions}
\AuthorContributions

\subsection*{Funding}
\FundingStatement

\subsection*{Competing interests}
\CompetingInterests

\subsection*{Ethics approval}
\EthicsApproval

\subsection*{Consent to participate}
\ConsentParticipation

\subsection*{Consent for publication}
\ConsentPublication

\subsection*{AI use declaration}
\AIUseDeclaration

\bibliographystyle{unsrtnat}
\bibliography{bib}

\end{document}